# Tuning Separator Chemistry: Improving Zn Anode Compatibility via Functionalized Chitin Nanofibers

Ibrahim Al Kathemi[a], Vishnu Arumughan[b], Mohamed Zbiri[c], Marcel Kröger[b], Ira Smal[b], Eero Kontturi[b], and Roza Bouchal[a*]

[a] *Department of Colloid Chemistry, Max-Planck Institute of Colloids and Interfaces, Am Mühlenberg 1, 14476 Potsdam, Germany*

[b] *Department of Bioproducts and Biosystems, Aalto University, FI-00076 Aalto, Finland*

[c] *Institut Laue-Langevin, 71 Avenue des Martyrs, Grenoble Cedex 9 38042, France*

corresponding author: roza.bouchal@mpikg.mpg.de

**Abstract**

Aqueous zinc (Zn) batteries (AZBs) face significant challenges due to the limited compatibility of Zn anodes with conventional separators, leading to dendrite growth, hydrogen evolution reaction (HER), and poor cycling stability. While separator design is crucial for optimizing battery performance, its potential remains underexplored. The commonly used glass fiber (GF) filters were not originally designed as battery separators. To address their limitations, nanochitin derived from waste shrimp shells was used to fabricate separators with varying concentrations of amine and carboxylic functional groups. This study investigates how the type and concentration of these groups influence the separator's properties and performance. In a mild acidic electrolyte that protonates the amine groups, the results showed that the density of both ammonium and carboxylic groups in the separators significantly affected water structure and ionic conductivity. Quasi-Elastic Neutron Scattering (QENS) revealed that low-functionalized chitin, particularly with only ammonium groups, promotes strongly bound water with restricted mobility, thereby enhancing Zn plating and stripping kinetics. These separators exhibit exceptional Zn stability over 2000 hours at low current densities (0.5 mA/cm$^2$), maintaining low overpotentials and stable polarization. Additionally, the full cell consisting of Zn‖$NaV_3O_8 \cdot 1.5H_2O$ showed a cycle life of over 2000 cycles at 2 A/g, demonstrating the compatibility of the nanochitin-based separators with low concentrations of functional surface groups. These results demonstrate the importance of a simple separator design for improving the overall performance of AZBs.

## 1. Introduction

The development of safe and reliable energy storage technologies is important because of the escalating global energy demands and the critical need to integrate renewable energy sources seamlessly. Aqueous zinc (Zn) batteries (AZBs) are an emerging technology with the capability to fulfil this demand and revolutionize the energy storage industry, due to their cost, safety, and scalability compared to lithium-ion technology[1]. However, a significant challenge facing AZB lies in its limited cycling stability. Overall, the issues associated with Zn anodes can be broadly categorized into hydrogen evolution reaction (HER), surface passivation from the heterogeneous deposition/dissolution of $Zn^{2+}$, corrosion, and dendrite formation[2]. These challenges impact each other, collectively compromising the reversibility and cycling performance of Zn anodes. Thus, the primary focus of AZB's research has shifted towards exploring materials and strategies aimed at reducing side reactions, improving electrochemical stability, and inhibiting dendrite growth[3–5].

The potential to enhance AZB's performance through separator design has often been overlooked[6]. Traditionally, the separator has been considered an inert component within the battery system, with the primary function being to effectively isolate the positive and negative electrodes [7,8], thereby preventing direct contact that could lead to short circuits[9]. However, the role of the separator is more important: it facilitates the transport of ions between the electrodes, either enhancing or impeding the movement of specific ions and molecules[10]. In addition, the separator is in direct contact with the electrodes, thereby influencing the interface chemistry[11]. Moreover, key characteristics such as porosity, thickness, zincophilicity, and electrolyte compatibility directly affect ion transport efficiency, Zn deposition behavior, and cycling stability[9,12]. Although glass fiber (GF) and cellulose (CLL) are widely used as filters in various applications, they are not well-suited for battery applications due to their inherent inability to address these requirements. For instance, the large and uneven pore size of GF and CLL separators compromises their mechanical properties and causes non-uniform $Zn^{2+}$ ion flux, promoting dendrite formation, aggregation, and potential short circuits in high-flux regions[13]. Moreover, these separators require a large amount of electrolyte to achieve complete wetting, resulting in excessive electrolyte consumption and a consequent low energy density[14]. Therefore, rational separator design offers a promising strategy to optimize these critical parameters, thereby improving battery performance and suppressing detrimental side reactions.

Anisotropic colloids derived from bioresources, such as nanocellulose and chitin nanofibers, are emerging as promising candidates for battery separators, particularly in AZBs. Their appeal stems from inherent advantages, including excellent mechanical properties, tunability of surface chemistry, and ease of processing[15,16]. The surface chemistry of these colloids governs their physicochemical behavior, influencing the water environment and its dynamics[17–20]. Additionally, the ions present in the electrolyte affect water uptake and interactions within these systems[21]. Therefore, the careful tuning of the surface chemistry of these anisotropic colloidal systems can significantly improve the overall AZBs' performance. There have been a few accounts validating nanochitin's effectiveness as a separator and electrolyte additive. However, these studies generally treat these bio-colloids as passive physical scaffolds rather than active chemical modulators in the system. A systematic investigation into how the distinct surface chemistries of these colloids modulate water activity and ionic conductivity is conspicuously absent from the literature. Addressing this gap is imperative, as these interfacial chemical interactions are the fundamental drivers that regulate local water structure, ensure electrolyte stability, and promote highly reversible Zn plating/stripping at the electrode interface.

This study systematically investigates how the identity and density of surface functional groups influence AZB performance, utilizing chitin nanofibers extracted from fisheries waste. We introduce carboxymethylation as a robust functionalization strategy to generate ampholytic chitin nanofibrils bearing carboxylate and amine groups. This approach was selected to circumvent the limitations of the common carboxylation method for polysaccharide nanoparticles, namely oxidation catalyzed by (2,2,6,6-tetramethylpiperidin-1-yl)oxyl (TEMPO), which is prone to unwanted side reactions between primary amines in the chitin and aldehyde intermediates during the oxidation[22]. Consequently, we were able to isolate and examine the specific roles of intrinsic amine groups versus mixed amine-carboxylate functionalities within the separator architectures in the AZB performance. The structural features and electrolyte dynamics of the resulting materials were probed using advanced characterization techniques, including Raman spectroscopy and quasi-elastic neutron scattering (QENS). The impact of these surface chemistries on the Zn anode stability was further assessed through comprehensive electrochemical and surface analyses. Finally, practical viability was validated in a full-cell configuration employing a vanadium-based cathode. These results demonstrate that the rational design of charged chitin nanofibers significantly enhances electrochemical performance, offering a pathway to more stable and efficient aqueous Zn ion batteries.

## 2. Results and discussion

### Preparation and structural properties

Chitin nanofibers, sourced from shrimp shells (Fig. 1a), were deacetylated to varying degrees to introduce primary amine groups. This process yielded two distinct samples: one with a primary amine content of 0.780 mmol/g (corresponding to a degree of deacetylation (DD) of 15.35%) and another with 1.74 µmol/g (DD of 32.94%). Following deacetylation, the fibers underwent surface-specific carboxymethylation to introduce carboxylic functionalities. Conductometric titration was used to quantify the incorporation of these carboxylate groups into chitin fibers with different degrees of deacetylation (Fig. S1 and S2). These steps resulted in a total of four different fibers with different surface chemistries: namely, fibers containing only an amine group at low and high contents ($NH_2$-Low and $NH_2$-High) and their carboxymethylated variants ($NH_2$/COOH-Low and $NH_2$/COOH-High). Note that the "Low" and "High" terms refer only to the amine functionality. The concentration of amine and carboxylate in these different fibers is given in Table S1 and Fig. 1c. The chemically modified fibers were mechanically disintegrated at an acidic pH to produce nanoscale building blocks for the separators. This acidic environment ensures the protonation of the primary amine ($NH_2$) groups present on the fiber surface, resulting in positively charged $NH_3^+$groups. This electrostatic repulsion generated by the protonated amine groups is the key mechanism providing the necessary colloidal stability for the resulting material in the dispersion.

As shown in the atomic force microscopy (AFM) micrographs (Fig. 1c), the resulting materials exhibited a typical fibrillar morphology. The average thickness of these nanofibers was determined to be in the range of 3-6 nm (Fig. 1d). A notable observation was that the fibers with low deacetylation degree (or low-$NH_2$ content) had a slightly higher average thickness compared to their high-deacetylated counterparts. This morphological difference is due to variations in electrostatic repulsion during the nano-fibrillation process at acidic pH, where highly deacetylated fibers would promote more effective lateral disassembly and yield thinner fibrils. However, this may also emerge from the "peeling" effect of the surface chains, which can happen due to the harsher deacetylation process. The surface charge characteristics of these nanofibrils were evaluated in different pH conditions, revealing that all these fibers are predominantly positively charged in pH < 6 due to protonation of amine groups (Fig. 1e-f). The $NH_2$/COOH-Low and $NH_2$/COOH-High exhibited a negative surface charge in high pH conditions, demonstrating ampholytic behavior, which is further representative of the successful chemical modifications.

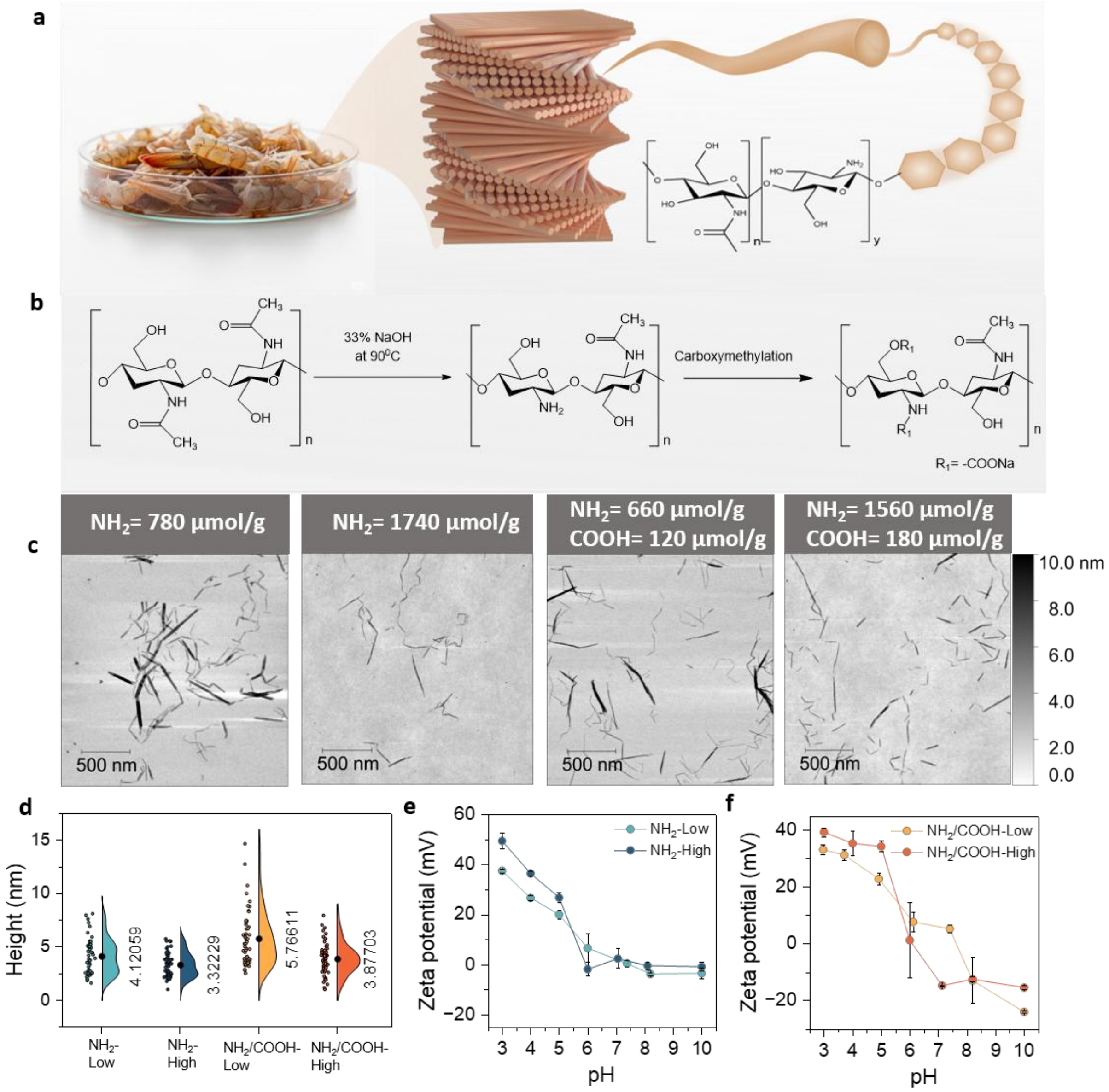

Fig. 1. Preparation and characterization of the chitin-based suspension. a) schematic overview of the recovered chitin. b) Chemical modification. c) Size distribution. d) Height distribution. e-f) Determined zeta potential at different pH values.

Separator membranes were fabricated from the nanofiber building blocks via a simple vacuum filtration followed by cold-press drying. This method produced freestanding membranes with a thickness of approximately 30 µm. This is significantly thinner than commercially available GF and CLL separators, which have thicknesses of 260 µm and 180 µm, respectively (Fig. S3). The commercial CLL separators used in this study were Whatman cellulose filter papers, manufactured from cotton liners treated to achieve an α-cellulose content of greater than 98%. The obtained α-cellulose was then processed by Büchner or Hirsch funnel filtration to produce the commercial CLL separator[23]. The reduced thickness of the chitin-based separators could shorten the ion migration distance, a key factor in enhancing ion transport properties[24]. In this study, a conventional 1 M Zn trifluoromethanesulfonate ($Zn(CF_3SO_3)_2$, $Zn(OTf)_2$) electrolyte was used with all separator types. A comparison of the physical properties of the dry and wetted separators is shown in the photographs in Fig. S3. The commercial GF and CLL separators were opaque and required at least 40 µl of electrolyte to be fully wetted and remained rigid after wetting (Fig. 2a). In contrast, the nanochitin-based separators needed less than 25 µl of electrolyte for complete wetting, which will contribute to higher volumetric energy densities[25]. Furthermore, the chitin-based separator membranes exhibited excellent mechanical properties, retaining their flexibility even after being saturated with electrolyte (Fig. S4 and Table S2). This characteristic is crucial for accommodating the volume changes that occur in the Zn anode during repeated plating and stripping cycles.[26]

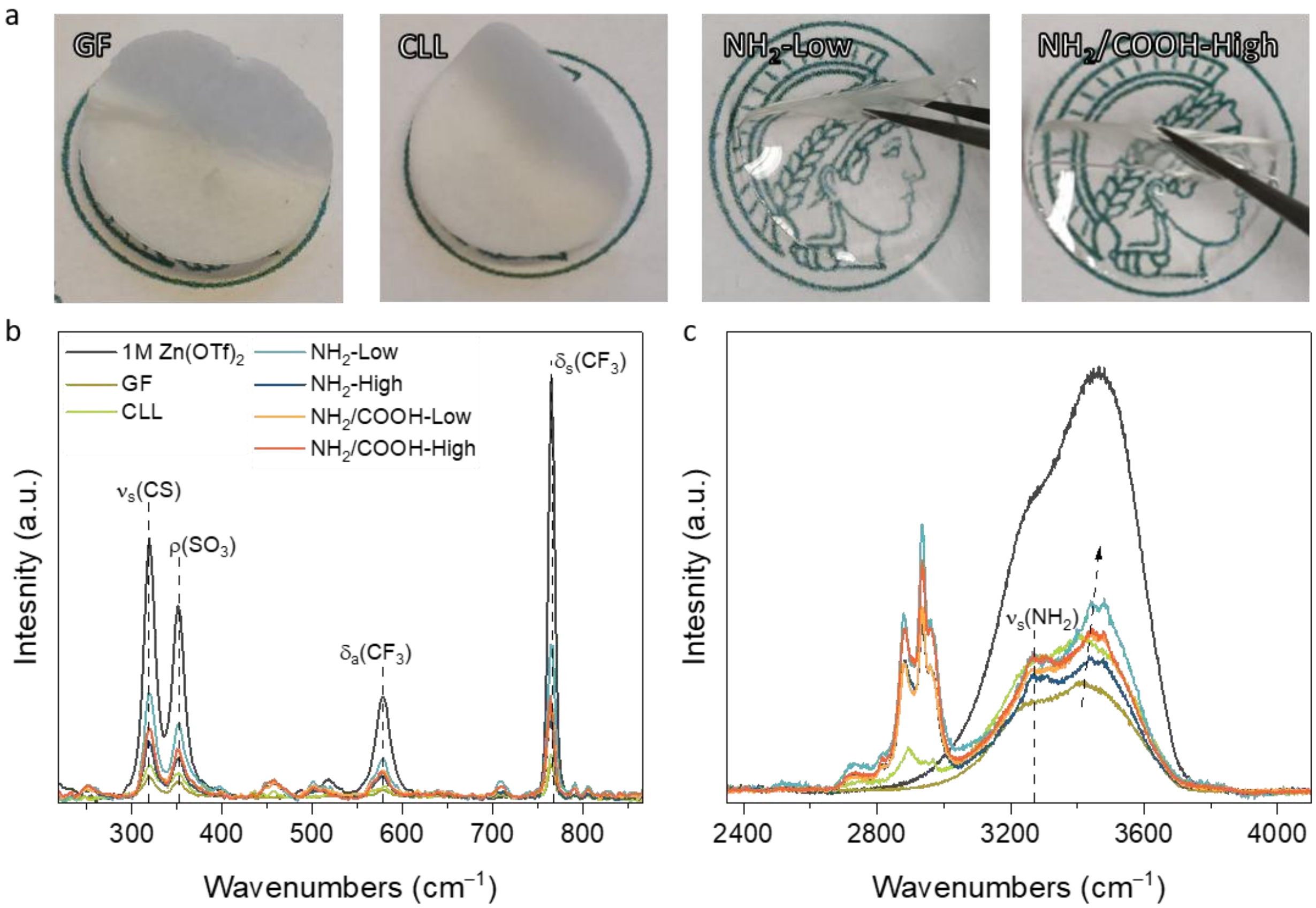

Fig. 2. Separator characterization. a) Flexibility of the separators after wetting. Raman spectra of the separators in regions b) 225 to 860 $cm^{-1}$ and c) 2390 to 4200 $cm^{-1}$.

To examine the microscale structural morphology, scanning electron microscopy (SEM) images of the pristine separators were captured. The SEM images (Fig. S5) showed that the GF separator consisted of cross-stacked amorphous silica fibers with varying diameters, creating large and irregular pores. In contrast, the CLL separator had a fibrous structure with low porosity because the fibers were closely adhered to each other. All chitin-based separators displayed no visible pores, indicating nanosized porosity in the structure. Although all chitin-based separators were similar in morphology, the $NH_2$/COOH-Low featured a rougher and more fibrous surface. Furthermore, Raman spectroscopy was employed to investigate the interaction between the chitin separators and the $Zn(OTf)_2$ electrolyte (Fig. 2b-c, Fig. S6-7). Spectra were collected for separators immersed in 1 M $Zn(OTf)_2$, and compared with those of the dried separators and the electrolyte solution. When comparing immersed chitin separators with the pristine separators, the characteristic amide I band in α-chitin was present in all spectra at 1628 - 1668 $cm^{-1}$ [27,28] (Fig. S6b, S7), with the first peak around 1628 $cm^{-1}$ assigned to stretching of the C=O groups hydrogen bonded both to NH and OH groups in the same chain and the second peak around 1648 $cm^{-1}$ to stretching of C=O groups hydrogen bonded to NH of the neighboring chain[29]. This indicated that the method of separator preparation did not affect the chitin backbone.

Furthermore, analysis of the OTf anion vibrations in the electrolyte and in all separators shows characteristic bands corresponding to the C-S symmetric stretching ($\nu_s$) at 319 $cm^{-1}$, $SO_3$ rocking (ρ) at 351 $cm^{-1}$, $CF_3$ asymmetric deformation ($\delta_a$) at 580 $cm^{-1}$, $CF_3$ symmetric deformation ($\delta_s$) at 766 $cm^{-1}$, $SO_3$ symmetric stretching ($\nu_s$) at 1032 $cm^{-1,}$ and $CF_3$ symmetric stretching at 1229 $cm^{-1}$ (Fig. S6-7)[30]. This indicates that the environment around the triflate ion was not altered upon contact with the chitin-based separators. However, when comparing the peak intensities of the OTf anion in the separators, in particular the $SO_3$ and $CF_3$ bands at 1032 $cm^{-1}$ and 765 $cm^{-1}$, respectively, the intensities in the chitin-based separators are higher than those in both CLL and GF separators. In addition, a very subtle downshift is observed for $SO_3$ and $CF_3$ bands to 1031 $cm^{-1}$ and 764 $cm^{-1}$, respectively, only in highly concentrated functionalized chitin ($NH_2$-High and $NH_2$/COOH-High). The increase in peak intensity may reflect higher electrolyte uptake in the chitin separators, particularly in the $NH_2$-Low sample, while the downshift might indicate stronger interactions or confinement within the chitin matrix. The NH and OH vibration modes of chitin in the 3000 - 3800 $cm^{-1}$ region are slightly broadened and increase in intensity when chitin is immersed in the electrolyte (Fig. S7). However, there is no clear correlation between the different chitin separators (Fig. 2c). Additionally, the OH asymmetric vibration of water around 3400 $cm^{-1}$ in both GF and CLL has shifted to higher wavenumbers in all nanochitin samples (Fig. S7). In addition, the peak intensity is higher in the low-concentration functionalized chitin, particularly for $NH_2$-Low. This latter is attributed to changes in the water environment, resulting in stronger water binding to the chitin nanofibers.

**Transport properties and dynamics**

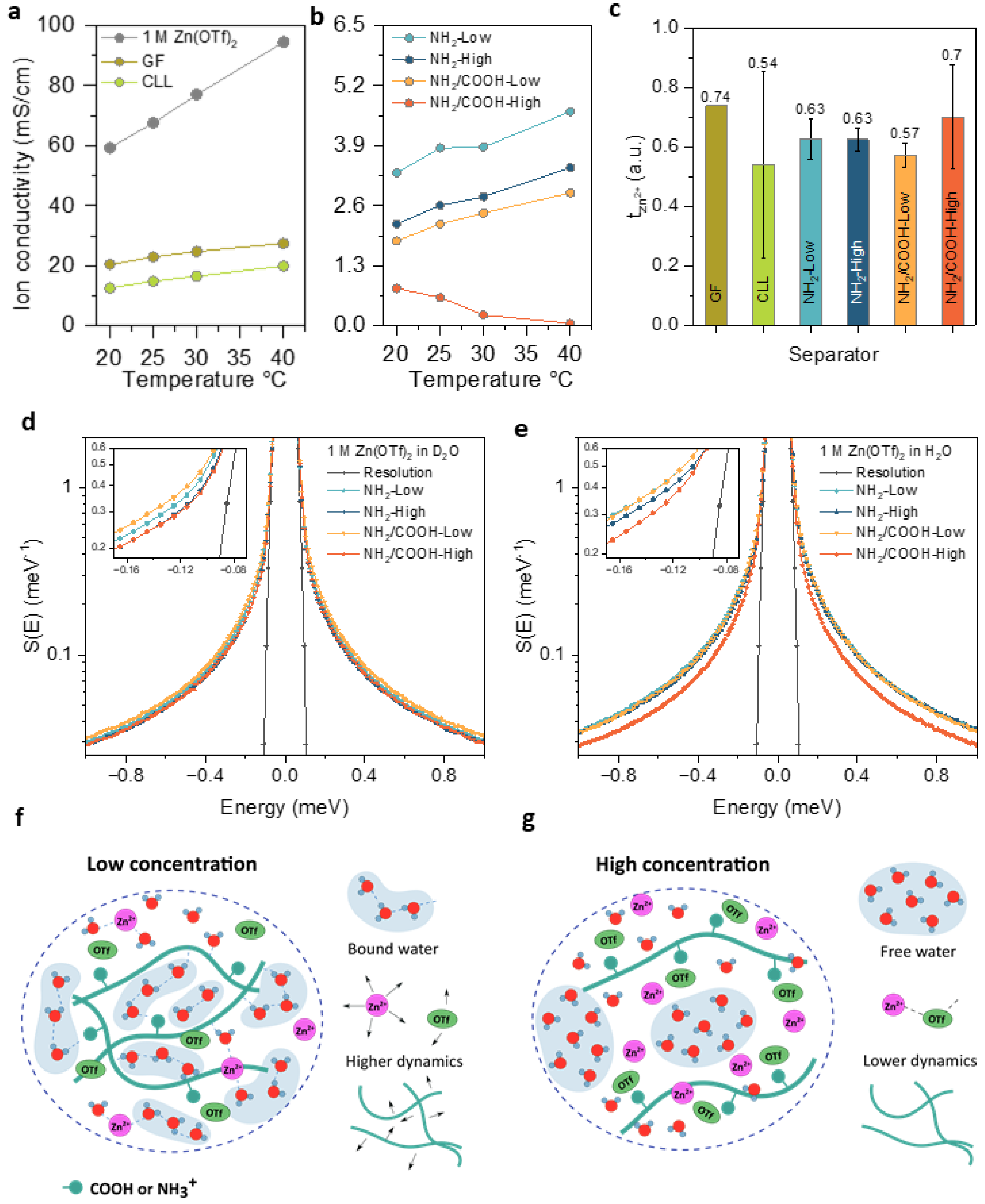

Fig. 3. Electrochemical properties and local environment of the chitin separators. a-b) Ionic conductivity (error bar of GF is removed because it was bigger than 1). c) Zn transference number. d-e) QENS spectra of the separators in 1 M $Zn(OTf)_2$ in $D_2O$ and $H_2O$, respectively. f-g) Schematic overview of chitin surface interactions for low and high charge concentration, respectively.

The ionic conductivity of the different separators was recorded from 20 to 40 °C, as pictured in Fig. 3a-b. GF and CLL showed high ionic conductivities across the entire temperature range. This was hypothesized to result from their higher porosity compared to the chitin-based separators (Fig. S5). In combination with the high ionic conductivity of 1 M $Zn(OTf)_2$ (68 mS/cm at 25 °C), the measured ion conductivity for the GF and CLL can be mainly attributed to the aqueous electrolyte[31]. The measured ionic conductivity for the chitin-based separators measured at 25 °C was 3.8 mS/cm for $NH_2$-low, 2.6 mS/cm for $NH_2$-High, 2.2 mS/cm for $NH_2$/COOH-Low, and 0.6 mS/cm for $NH_2$/COOH-High. These values are comparable to what is found for bio-based separators in literature, for example, carboxymethyl reinforced cellulose nanofibers (1.34 mS/cm)[32], chitosan grafted bacterial cellulose (2.9 mS/cm)[33], and polydopamine-modified cellulose (1.54 mS/cm)[34]. Moreover, the ionic conductivity of all the chitin separators increases with temperature except with $NH_2$/COOH-High, which approaches 0 mS/cm at 40 °C. The latter can be attributed to the formation of ion pairs, as previously observed for highly concentrated electrolytes[35]. Overall, the results show a clear impact of the functional groups and their density on the ionic conductivity, with $NH_2$-only samples exhibiting the highest value.

Additionally, the transference number ($t_{Zn2+}$) was determined using the Bruce-Vincent method by applying a 30 mV potential for 30 minutes (Fig. S8)[28]. A higher $Zn^{2+}$ transference number helps reduce the cation concentration gradient and promotes a more uniform $Zn^{2+}$ distribution, which is essential for preventing dendritic growth[36]. Fig. 3c suggested that the $NH_2$-low, $NH_2$-High, and $NH_2$/COOH-Low had similar $t_{Zn2+}$ of 0.63, 0.63, and 0.57, respectively. $NH_2$/COOH-high sample displayed a slightly higher value of 0.7, but the error bar was quite high, as for GF and CLL, with a $t_{Zn2+}$ of 0.74 and 0.54, respectively. The measured values of the nanochitin samples are slightly higher than the reported $t_{Zn2+}$ in the literature for chitin-based separators, such as crosslinked chitin fibrils in gel electrolyte (0.468)[37] and pan-milled TEMPO-oxidized chitin nanofibers (0.49)[38]. However, the similarities in Zn transference numbers across the different chitin separators indicate that the higher ionic conductivity observed in the $NH_2$-only and low-concentration $NH_2$/COOH-chitin samples may be related either to enhanced anion mobility[39,40] or to the limited accuracy of this method in concentrated 1 M electrolytes, as indicated by the large error bars[41]. Therefore, it is difficult to properly assess the impact of the functional groups on the $Zn^{2+}$ mobility.

Due to the large incoherent neutron scattering cross-section of hydrogen, QENS provides insights into the microscopic single-particle dynamics on the nanometer length scale and the picosecond time scale. Additionally, QENS offers a unique feature to tune the hydrogen signal by using the deuteration technique for contrast variation purposes[42]. In this context, to gain better insights into the role of amine and carboxylic groups on the overall dynamics of the nanochitin based separators, QENS measurements were performed on the different chitin separators under four environments: (i) dried (pristine) chitins, probing intrinsic chitin dynamics; (ii) chitins in pure $H_2O$, probing the effect of the electrolyte-free hydration; (iii) chitins in a 1 M $Zn(OTf)_2$ electrolyte with deuterium oxide ($D_2O$) as solvent, suppressing water scattering to highlight chitin only responses to the electrolyte uptake; and (iv) chitins in a 1 M $Zn(OTf)_2$ electrolytes with $H_2O$ as solvent, where the hydration-water signal dominates and directly reflects ionic mobility.

QENS spectra of dried (pristine) chitins report their intrinsic dynamical behavior with respect to each other (stiff, flexible, etc). The QENS spectra of the dried chitins with either low or high -$NH_2$/COOH samples are narrower than the chitins with only the -$NH_2$ group (Fig. S9). This indicates that the -COOH group has an immobilization effect which, under hydration, could act as ion traps that would reduce ionic dynamics, potentially reflecting a reduction of $Zn^{2+}$ conductivity. The nanochitin with primarily a low concentration of -$NH_2$ remains relatively mobile (broader QENS spectra), in agreement with higher ionic conductivity measurements.

QENS spectra of chitins in pure $H_2O$ are expected to be dominated by hydration water dynamics, and thus the spectra would mainly reflect the mobility of the hydration spheres (Fig. S9). The QENS spectra in this case show that the low-$NH_3^+$ concentration separators have narrower signals than their high-$NH_3^+$ concentration counterparts. The $NH_2$/COOH-High separator becomes mostly plasticized with $H_2O$. This indicates that the chitin-$H_2O$ interaction landscape changes strongly depending on the $NH_2$ and/or COOH groups, with low COOH concentration leading to a stiff chitin in the dry state, while high COOH concentration leads to a plasticized chitin in $H_2O$. The latter may originate from a swelling due to the formation of free water clusters with higher mobility.

For chitin-based separators immersed in deuterated electrolyte, the QENS captures how the chitin dynamics respond to the presence of salt (Fig. 3d). The resulting QENS spectra reveal that the deuterated electrolyte induces plasticization in $NH_2$-Low and $NH_2$/COOH-Low chitins, as evident by the spectral broadening associated with increased molecular mobility, indicating favored ion transport. In contrast, $NH_2$-High and $NH_2$/COOH-High samples exhibited the opposite trend, showing restricted dynamics indicative of structural stiffness and potential ion trapping or ion pairing. The systems of chitins in pure $H_2O$ and in deuterated $Zn(OTf)_2$ are mainly relevant for understanding how chitins intrinsically interact (or bind) with water and ions, respectively. However, the case of chitins in the $Zn(OTf)_2$ electrolyte in water provides the most direct observation of $Zn^{2+}$ and OTf ion dynamics, and whether the electrolyte acts to plasticize or reorganize the chitin structural network (Fig. 3e). In the context of protonated electrolyte, the QENS spectra reveal an additional behavior of the chitins, namely that broader spectra were observed for the $NH_2$-Low and $NH_2$/COOH-Low samples, indicating enhanced dynamics and higher ionic conductivity, whereas the narrower spectra of the $NH_2$-High and $NH_2$/COOH-High samples reflect ion trapping and reduced mobility. Overall, the QENS results clearly demonstrate that a lower charge density within the chitin matrix promotes higher overall dynamics and ionic mobility.

By combining all the data, a hypothesis can be proposed. The 1 M $Zn(OTf)_2$ electrolyte exhibits a pH of approximately 5, under which the amine groups in the separators are largely protonated (-$NH_3^+$), imparting a net positive surface charge. Under these mild-acidic conditions, the low-concentration separator may act as a single-ion conductor, wherein the anionic OTf species experience restricted mobility or partial immobilization through electrostatic interactions with the protonated -$NH_3^+$ groups. Raman spectral shifts further corroborate this behavior, showing changes predominantly associated with the OTf environment, indicative of altered anion coordination within the separator matrix. Complementary QENS spectra of low-concentration chitin separators reflect enhanced ionic mobility and confined water molecules (Fig. 3f). In contrast, the QENS observations for high-concentration chitins suggest restricted ion dynamics, consistent with the formation of ion aggregates or ion pairing with $Zn^{2+}$ and OTf ions (Fig. 3g), particularly with the $NH_2$/COOH-high sample. These results are consistent with the ionic conductivity measurements, except for the opposite trend observed in the QENS data, where the $NH_2$/COOH-Low sample exhibits slightly higher dynamics than the $NH_2$-High sample. This behavior can be attributed to contributions from water dynamics, which are not captured in ionic conductivity measurements. The QENS data suggest that water confinement is enhanced at lower chitin concentrations, particularly in the $NH_2$-Low sample. Conversely, the higher water mobility observed in the high-concentration chitin samples, especially $NH_2$/COOH-High, may promote parasitic reactions and compromise Zn anode stability.

**Zn metal stability**

Continuing with the electrochemical analysis, potentiostatic electrochemical impedance spectroscopy (PEIS) tests were carried out at various temperatures to examine the desolvation energy ($E_{des}$) of $Zn^{2+}$ ions. From Fig. S10 and Fig. S11, it could be concluded that the separators CLL, $NH_2$-Low, $NH_2$-High, and $NH_2$/COOH-Low had similar calculated $E_{des}$ (Eq. S3) of 30, 29, 30, and 31 kJ/mol, respectively. The GF and $NH_2$/COOH-High had slightly Lower $E_{des}$ of 24 and 22 kJ/mol, respectively. These results indicated that the chitin-based separators have almost similar $E_{des}$ for $Zn^{2+}$ ions, compared to commercial CLL and GF. Furthermore, the results obtained are consistent with the literature. Wang et al. reported a chitin-based separator composed of both crab and shrimp chitin with an $E_{des}$ of 30.0 kJ/mol[38], while Lin et al. proposed a chitin-based separator derived from crab chitin, which exhibited an $E_{des}$ of 27.2 kJ/mol[43].

Cyclic voltammetry (CV) curves for the Zn‖Zn cells are shown in Fig. S12. The GF separator exhibits the highest current density, indicating faster Zn oxidation and reduction. However, the cell cycled for only a short time, fewer than five cycles. In contrast, the $NH_2$-Low, $NH_2$-High, and $NH_2$/COOH-Low samples exhibit broader peaks, suggesting altered redox reactions and plating/stripping behavior, likely influenced by the positively charged chitin nanofibers. The CLL and $NH_2$/COOH-High samples show the lowest current densities, reflecting slower Zn kinetics. Additionally, the chitin-based electrolytes showed lower nucleation overpotential (inset Fig. S12), and the $NH_2$-low and $NH_2$/COOH-Low showed the highest anodic and cathodic peaks besides GF. This indicated their ability to effectively reduce the deposition barrier of $Zn^{2+}$ ions[36]. To investigate Zn nucleation and growth behaviors, current transients were analyzed by chronoamperometry, which can indicate changes in nucleation processes and surface morphologies [44]. As shown in Fig. S13, applying a constant potential of -150 mV led to a continuously increasing current for the GF, CLL, and $NH_2$/COOH-High samples, indicating rapid 2D planar diffusion and coarse Zn deposition, conditions that may promote Zn dendrite formation[45]. In contrast, the $NH_2$/COOH-Low sample exhibited a stabilized current after approximately 3.5 minutes, suggesting an initial 2D diffusion regime followed by a transition to 3D diffusion. Meanwhile, the $NH_2$-Low and $NH_2$-High samples showed a rapid shift to 3D diffusion, indicating uniform $Zn^{2+}$ deposition across the electrode surface as $Zn^{2+}$ concentration increased. These results demonstrate that the nanochitin separator with a low concentration of functional groups promotes a more uniform Zn electrodeposition than the currently commercialized GF and CLL separators[38,43].

The long-term cycling stability of Zn anodes was assessed in symmetric ZnIIZn cells. First, a low current density of 0.5 mA/cm$^2$ and a capacity of 0.4 mAh/cm$^2$ were used (Fig. S14). The GF and CLL were underperforming, with a maximum cycle life of 140 hours. In contrast, the $NH_2$-Low and $NH_2$/COOH-Low separators both cycled for over 2000 hours, with $NH_2$-Low showing less potential polarization, indicating better plating and stripping behavior. The $NH_2$-High showed stable cycling for 1200 hours until high polarization was visible, while the $NH_2$/COOH-High cycled for only 700 hours. At a higher current density and capacity of 1 mA/cm$^2$ and 1 mAh/cm$^2$, respectively (Fig. 4a), the GF, CLL, $NH_2$-High, and $NH_2$/COOH-High cells operated for approximately 30 to 50 hours, whereas the $NH_2$-Low and $NH_2$/COOH-Low cells cycled for over 900 hours under the same conditions. The poor performance of the highly functionalized nanochitin at higher current densities is directly related to its reduced ionic dynamics. This suggested that a higher degree of deacetylation and carboxymethylation hinders the Zn stripping and plating mechanism. To evaluate the $NH_2$-Low and $NH_2$/COOH-Low under more realistic conditions, relevant to stationary storage applications, which typically require cycling rates between C/4 and C/6[46], the current density was kept at 1 mA/cm$^2$ while the capacity was increased to 5 mAh/cm$^2$. This corresponded to a C-rate of C/5, equivalent to a 5-hour charge/discharge condition. Both $NH_2$-Low and $NH_2$/COOH-Low showed a low overpotential for over 700 and 500 hours, respectively (Fig. S15), indicating the high compatibility of the chitin fibril-based separators and the Zn anode. Maintaining the same high capacity, the cells were also tested at a current density of 5 mA/cm$^2$ (Fig. S16), corresponding to a 1C rate. Even under these conditions, the $NH_2$-Low and $NH_2$/COOH-Low separators exhibited lifetimes exceeding 350 hours (175 cycles), compared with less than 20 cycles for GF and CLL. Symmetric Zn cell tests confirmed the excellent compatibility of $NH_2$-Low and $NH_2$/COOH-Low with the Zn anode. The -COOH groups showed minimal influence on Zn stability, whereas higher charge density negatively affected it.

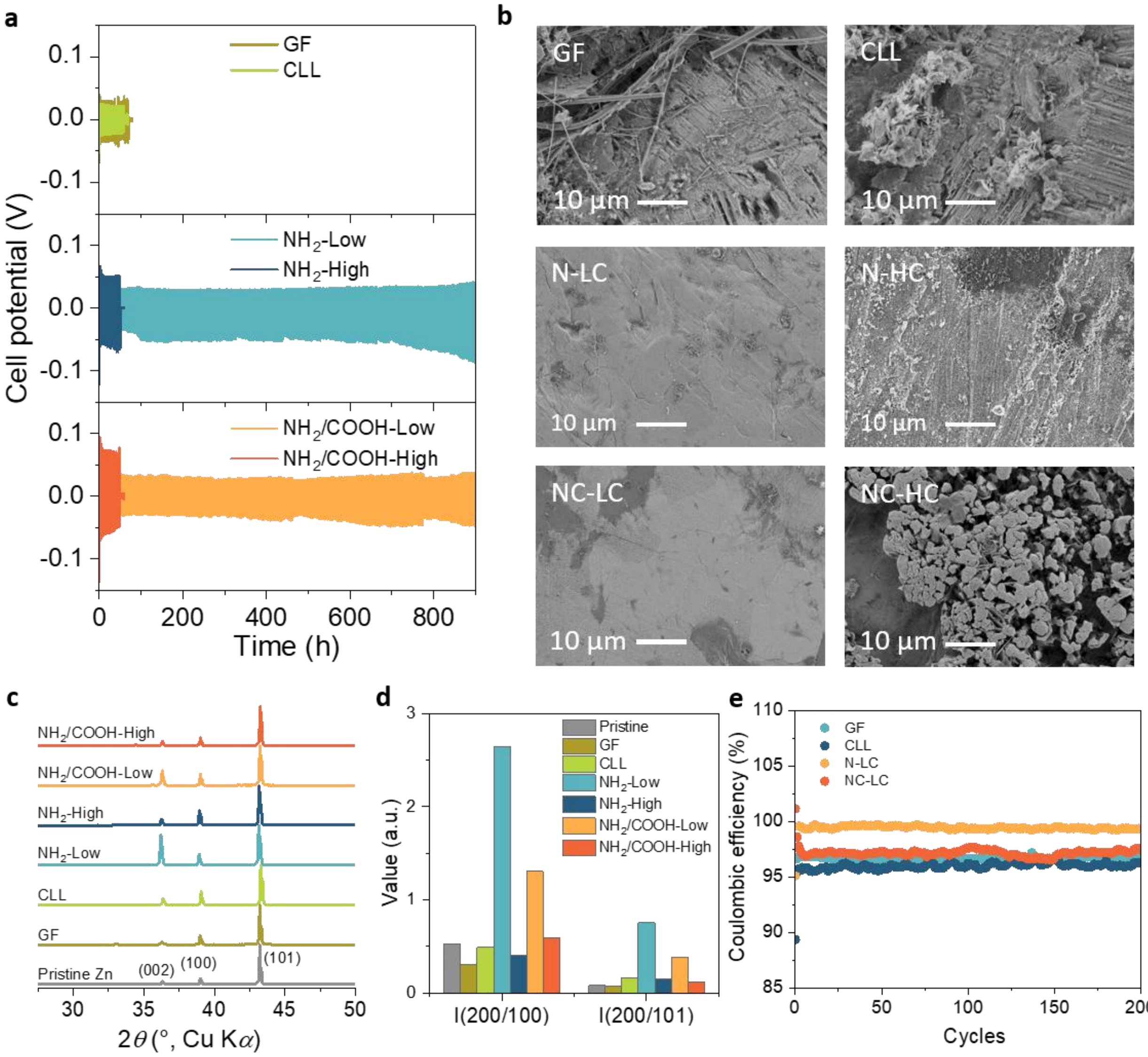

Fig. 4. Zinc anode performance. a) Long-term stability with 1 mA/cm2 and 1 mAh/cm2. b) SEM images of the recovered Zn anodes. c) XRD spectra of the recovered Zn anodes. d) Peak intensity comparison of the XRD spectra. e) Determined coulombic efficiency.

To further understand the plating and stripping nature, both the chitin separators and the Zn surfaces were examined after cycling. Fig. S17 displays the plating/stripping curves of 20 cycles of the symmetric Zn‖Zn cell utilizing an applied current density and capacity of 1 $mA/cm^2$ and 1 $mAh/cm^2$, respectively. First, the surface of the separators after cycling was examined by SEM, as shown in Fig. S18. For the GF and CLL, it is evident that Zn is deposited on the fibers themselves, leading to a shorter lifetime. The same observation was noted with the $NH_2$-High separator and, to a lesser extent, with the $NH_2$/COOH-High separator, explaining their short lifetimes. However, the SEM images indicated that $NH_2$/COOH-Low had minor cracks, whereas $NH_2$-Low showed only small local damage, visible only at high magnification.

Subsequently, the SEM images of the recovered Zn foils are depicted in Fig. 4b and S19. For the Zn metal in contact with GF, CLL, $NH_2$-High, and $NH_2$/COOH-High, signs of corrosion and dendrite formation were observed. The Zn anode paired with $NH_2$/COOH-Low showed some dark spots, whereas the one in contact with $NH_2$-Low exhibited a smooth surface. This is further supported by the X-ray diffraction (XRD) spectra depicted in Fig. 4c. In the XRD spectra of Zn anodes, the (002), (100), and (101) peaks represent the main crystal planes of hexagonal Zn deposition. A strong (002) peak indicates that the Zn grows mainly along the basal plane, which helps prevent dendrite formation and improves battery stability. It was visible that the ratio of the (002) peak to the (100) and (101) peaks in low-charged samples was higher in comparison to the other separators (Fig. 4d), suggesting that $Zn^{2+}$ prefers to plate in the more energetically favorable (002) plane when $NH_2$-Low and $NH_2$/COOH-Low were utilized.

X-ray photoelectron spectroscopy (XPS) was taken from the recovered Zn anodes after 20 cycles and is shown in Fig. S20 and S21. The S 2p spectra could be deconvoluted into three doublets. The first is attributed to $*CF_3SO_3$ at 169.4 eV in the $NH_2$-Low spectrum, which shifts to a lower binding energy of 168.6 eV in $NH_2$-High, $NH_2$/COOH-Low, and $NH_2$/COOH-High samples. The peak attributed to ZnS was visible in all recovered Zn anodes at 163.1 eV, with the low-concentration separators indicating a higher presence of ZnS on the surface due to the higher intensity of the peaks[47–49]. $S^{-2}$ was also detected at around 160.6 eV for all separators. These are characteristic decomposition products of OTf-containing electrolytes, indicating the formation of a solid electrode interphase (SEI) layer for all separators[50–52]. The C 1s spectra also showed typical decomposition products of $Zn(OTf)_2$ electrolyte, such as C-O at 286.5 eV and $ZnCO_3$ at 289.2 eV for the recovered Zn anode cycled with the $NH_2$-High, $NH_2$/COOH-Low, and $NH_2$/COOH-High separators, with $ZnCO_3$ being a product when sufficient reduction of Zn-OTf complexes occurs[53]. However, the $NH_2$-Low separator indicated incomplete decomposition products of $Zn(OTf)_2$ electrolyte, namely C-H at 285.7 eV, C=O at 288.3 eV, and *CF at 290.3 eV[54], possibly indicating that fewer $OTf^-$ ions reached the Zn anode surface to decompose. The formation of C-H and C=O in the $NH_2$-Low was attributed to the decomposition of chitin instead, as previously reported in the literature[55,56]. The detection of R-$NH_2$ amine groups around 400 eV in the N 1s spectra indicates that the chitin-based separators also decomposed to form an SEI layer. The different detected decomposition products of $Zn(OTf)_2$ electrolyte in the $NH_2$-Low sample might indicate that the $NH_2$-Low sample follows a different decomposition reaction than the other three separators.

The O 1s spectra of $NH_2$-High, $NH_2$/COOH-Low, and $NH_2$/COOH-High are similar, with peaks of ZnO at 529.9 eV, $ZnCO_3$ at 531.7 eV, and O-S/C=O at 533.2 eV. While the formation of $ZnCO_3$ and O-S/C=O is from decomposition of the $Zn(OTf)_2$ electrolyte, the ZnO is present due to the decomposition of water molecules on the Zn anode surface[51,57,58]. Again, the spectra of $NH_2$-Low had a different spectrum, with the detection of $Zn(OH)_2$ at 530.9 eV instead of ZnO. Taking into consideration that the formation of ZnO is a two-step reaction, where $Zn(OH)_2$ should first form by decomposition of the water molecules, then reach saturation concentrations before further reacting to ZnO[59], the absence of ZnO in the $NH_2$-Low sample could prove that the SEI layer in this sample does follow a different decomposition route, one which combines the decomposition products of the electrolyte and separator to form a more robust SEI that blocks water molecules from reaching the surface and therefore inhibits the formation of unwanted side reactions such as the formation of ZnO. Additionally, the F 1s spectra of all separators revealed clear products of the reduction of $Zn(OTf)_2$ electrolyte, namely $ZnF_2$ at 684.5 eV and $*CF_3$ at 688.5 eV. The formation of $ZnF_2$ in the SEI is known to allow $Zn^{2+}$ to diffuse through the SEI while blocking the water molecules from reaching the surface, inhibiting parasitic reactions and dendrite formation[53].

Nevertheless, these results confirm the good compatibility of $NH_2$-Low with the Zn anode. Therefore, the low-charge chitin-based separators, $NH_2$-Low and $NH_2$/COOH-Low, were selected for further characterization. Coulombic efficiency (CE) tests using $NH_2$-Low and $NH_2$/COOH-Low in a Zn-Cu cell were performed and compared with GF and CLL (Fig. 4e and Fig. S22). $NH_2$/COOH-Low, GF, and CLL had close CE of 97, 96, and 95% average over 200 cycles, while $NH_2$-Low had over 99% efficiency. In summary, the $NH_2$-Low separator exhibited better overall performance for Zn anode stability, followed by $NH_2$/COOH-Low. This is likely because the uncharged carboxylic group in $NH_2$/COOH-Low does not significantly affect the $Zn^{2+}$ kinetics or interface compatibility with the Zn anode.

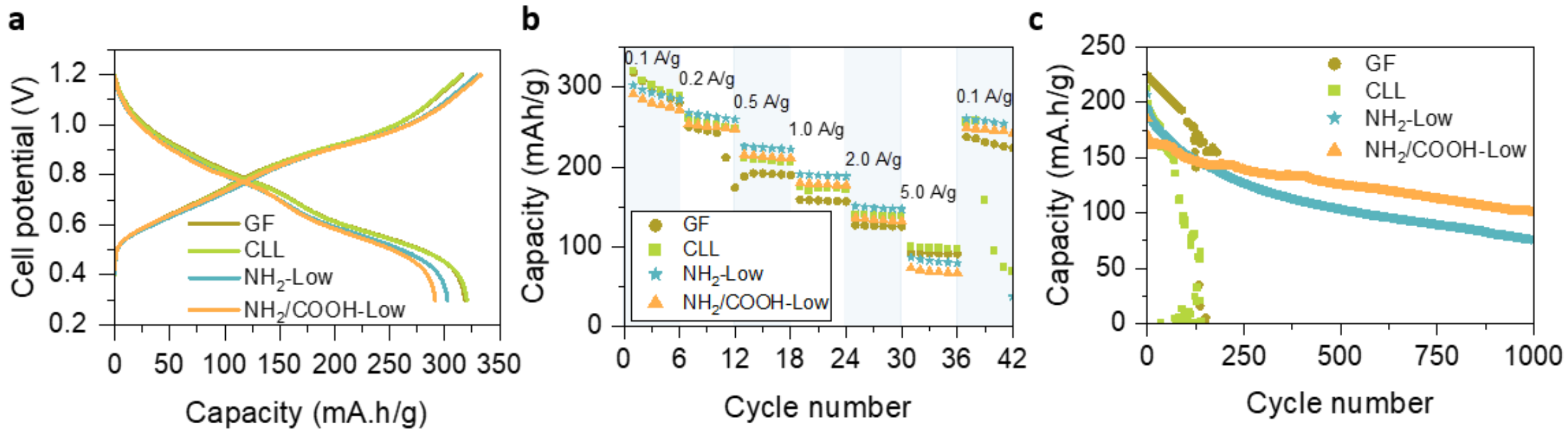

Fig. 5. Electrochemical performance of the $NaV_3O_8 \cdot 1.5H_2O$ cathode. a) GCD curves of the second cycle at 0.1 A/g. b) Rate capability. c) Long-term stability at 2 A/g.

Finally, as a showcase example, the low concentration samples were further evaluated against GF and CLL in full cells comprising a Zn anode and a NaVO-based cathode. The NaVO material was synthesised using the same procedure described in our previous publication[60], following the method of Wan et al.[61]. The galvanostatic charge/discharge profiles (Fig. 5a) for all separators were similar, with two plateaus during charge and discharge, indicating a two-step (de-)intercalation mechanism for VO cathodes. Different current densities were applied to test the rate capabilities of the separators. The capacities are shown in Fig. 5b, and the corresponding CE is depicted in Fig. S23a. The capacities of each separator at each current density were comparable, with slightly higher capacities in the NH2-Low sample. Overall, the stability of $NH_2$-Low and $NH_2$/COOH-Low outperformed GF and CLL. The GF showed instability at lower current densities, as evidenced by a capacity drop from cycle 10 to 12 at 0.2 A/g and a CE above 100%, indicating unwanted side reactions. As for the CLL, it did not recover its initial capacity when switching from a high current density (5 A/g) to the initial current density of 0.1 A/g. While the $NH_2$-Low and $NH_2$/COOH-Low showed more stable cycling and recovered most of their initial capacities, however display low CE at low current densities of 0.1 and 02 A/g, indicating possible side reactions that are inhibited at higher current densities. The long-term stability was evaluated by applying 2 A/g (Fig. 5c and Fig. S23b). The lifetimes of GF and CLL were around 75 cycles, while $NH_2$-Low and $NH_2$/COOH-Low cycled for over 1000 cycles. The CE of $NH_2$-Low and $NH_2$/COOH-Low reached 99.9% and 99.5%, respectively. However, $NH_2$-Low had a lower capacity retention of 38.7% compared to $NH_2$/COOH-Low's 61.5%. Although capacity fading remained a key challenge, the chitin nanofiber-based separators still demonstrated compatibility and improved kinetics compared to industrially available separators for AZIBs.

## 3. Conclusion

Waste biomass-based chitin nanofiber separators with different functional groups were developed for AZIBs. Through deacetylation and carboxymethylation, four types of chitin-based separators were prepared ($NH_2$-Low, $NH_2$-High, $NH_2$/COOH-Low, and $NH_2$/COOH-High). The separators exhibit excellent wettability, flexibility, and mechanical stability in contact with the electrolyte. QENS data reveal that both the type and density of functional groups in the chitin nanofibers strongly influence the local water environment and ionic dynamics. In particular, reduced water mobility is observed in the low-concentration separators ($NH_2$-Low and $NH_2$/COOH-Low) compared with the high-concentration counterparts. These findings correlate with ionic conductivity measurements, which are higher in the functionalized amine chains and low-concentration carboxylated nanofibers, providing a mechanistic explanation for the enhanced cyclability of the Zn anode observed with low-functionalized chitin-based separators. At 0.5 and 1 mA/cm$^2$ current densities, the $NH_2$-Low and $NH_2$/COOH-Low indicated stable cycling for over 2000 hours and 850 hours, respectively. Even at a low C-rate of C/5, the $NH_2$-Low and $NH_2$/COOH-Low exhibited a long cycle life of over 700 and 500 hours, respectively. The SEM and XRD further validated the compatibility of $NH_2$-Low and $NH_2$/COOH-Low with the Zn anode, showing a smooth surface and favored planar Zn deposition, respectively. Lastly, the $NH_2$-Low and $NH_2$/COOH-Low were tested in a full cell utilizing a NaVO-based cathode. Though capacity retention proved a bottleneck for the proposed separators, the rate capability and long-term cycling over 1000 cycles indicated the compatibility of $NH_2$-Low and $NH_2$/COOH-Low with NaVO-based cathodes. This study demonstrates that tuning the surface chemistry of chitin-based separators can effectively modulate the ion and water environments within the separators, thereby deepening our understanding of AZIBs and offering a sustainable, reliable alternative for next-generation separators.

## 5. Conflicts of interest

The authors declare no conflict of interest.

## 6. Acknowledgements

This work was supported by the Max Planck Society. We acknowledge the Institute Laue-Langevin (ILL) facility, Grenoble, France, for providing beam time on the IN5 spectrometer. R. B. acknowledges funding by the European Union's Framework Program for Research and Innovation Horizon 2020 (2014–2021) under the Marie Skłodowska-Curie Grant Agreement No. 101032227. V. A. and E. K. acknowledge Business Finland (IMD1, decision 613/31/2023) and the Research Council of Finland's Flagship Programme (Competence Center for Materials Bioeconomy, FinnCERES) for the infrastructure.

*Electronic Supplementary Information (ESI)*

# Tuning Separator Chemistry: Improving Zn Anode Compatibility via Functionalized Chitin Nanofibers

Ibrahim Al Kathemi[a], Vishnu Arumughan[b], Mohamed Zbiri[c], Marcel Kröger[b], Ira Smal[b], Eero Kontturi[b], and Roza Bouchal[a*]

[a] *Department of Colloid Chemistry, Max-Planck Institute of Colloids and Interfaces, Am Mühlenberg 1, 14476 Potsdam, Germany*

[b] *Department of Bioproducts and Biosystems, Aalto University, FI-00076 Aalto, Finland*

[c] *Institut Laue-Langevin, 71 Avenue des Martyrs, Grenoble Cedex 9 38042, France*

corresponding author: roza.bouchal@mpikg.mpg.de

## Contents

## Materials

Chitin (poly-(1-4)-β-N-acetyl-D-glucosamine) coarse flakes from shrimp shell were purchased from Sigma Aldrich. Sodium borohydride ($NaBH_4$) (Sigma Aldrich), Sodium hydroxide pellets (VMR chemicals), Monochloroacetic acid (Sigma Aldrich), Isopropanol (VMR Chemicals), Methanol (VMR chemicals), Ethanol 99% (Anora Industrial), Glacial acetic acid (CH3COOH, >=99.7%, CAS 64-19-7, analytical reagent grade, Fisher Scientific), Ammonia solution (NH4OH, 25%, for analysis, CAS 1336-21-6 ISO Reag. PH Eur, Supelco (Merck)), Hydrogen chloride solution (HCl, 0.1 M, CAS 7647-01-0, Titripur Reag., Merck) were used without further purification. Milli-Q water (18.2 MΩ cm resistivity) was degassed for 1 hour before use in titrations. Sodium hydroxide solution (NaOH, 1 M, CAS 1310-73-2, Titripur Reag., Merck) was diluted to a concentration of 0.1 M with degassed Milli-Q water and used for conductometric titrations.

Zinc trifluoromethanesulfonate ($Zn(OTf)_2$) was purchased from Sigma Aldrich. Demi water was utilised to obtain the 1 M $Zn(OTf)_2$ electrolyte. Whatman glass microfiber grade GF/C filter discs with 1.2 µm pore size and Whatman quantitative cellulose filter paper of grade 44 were purchased from Cytiva. Copper foil (99.95%, 10 µm) was purchased from MTI. Zinc foil (99.95%, 100 µm) was purchased from Goodfellow. Alumina polishing suspension (0.05 µm) from Allied High Tech Products Inc. was used to clean the zinc foils, followed by sonicating for 5 minutes in acetone.

## Separator synthesis

**Partial deacetylation of chitin**

The partial deacetylation of the chitin fibers was carried out using a method described elsewhere[1,2]. Briefly, A dry weight of 40g of chitin fibers was treated with 1 liter of 33% NaOH solution at 90 °C. The treatment time was varied to produce chitin fibers with different degrees of deacetylation (4 hours and 24 hours). To minimize depolymerization, 1.2 g of sodium borohydride was added. After the reaction, the product was filtered and thoroughly washed to remove any excess ions.

**Carboxymethylation of partially deacetylated chitin**

Carboxymethylation of partially deacetylated chitin fibers was carried out according to a method adapted from the carboxymethylation of cellulose[3]. A 15g of partially deacetylated fibers were blended using a cold disintegrator, and solvent exchanged to ethanol by washing them in ethanol three times with an intermediate filtration step. The fibers were then impregnated with a solution of 1.35g of monochloroacetic acid in 65 ml of isopropanol. These fibers were then added in portions to a solution of 2.3g of NaOH in 75 ml of methanol mixed with 300 ml of isopropanol that had been heated to just below its boiling temperature in a reaction vessel fitted with a condenser. This carboxymethylation reaction was allowed to continue for 1 hour. Following the carboxymethylation step, the fibers were filtered and washed first three times with 1 liter of deionized water, then with 200 ml of HCl (0.1M), and finally with deionized water until the filtrate was neutral.

**Disassembly of chitins into nanofibrils**

A 0.3 wt% suspension of chitin fibers was adjusted to pH 3 using 0.1 M HCl. Following the pH adjustment, the suspension was mechanically disintegrated using a high-speed blender (T-25 Ultra-Turrax Digital Homogenizer, IKA, Germany) at 10,000 rpm for 5 minutes. Subsequently, the suspension underwent microfluidization (M-110P, Microfluidics Inc., Newton, MA, USA) at a pressure of 1500 bar for one pass. The fluidized suspension was then centrifuged at 4500 rpm for 15 minutes using a Thermo Fisher Scientific Jouan GR-4.22 centrifuge. Finally, the colloidally stable supernatant was collected and stored for further use.

**Preparation of separators from nanochitin building blocks**

A 20 g portion of the ChNF suspension, corresponding to 0.033 g of dry ChNF, was adjusted to pH 6.2 using 0.1 M NaOH. Subsequently, 10 mL of ethanol (99%) was added to the suspension. The mixture was then probe-sonicated in an ice bath using a Branson S-450 sonicator unit at 10% amplitude with 30-second pulses. Following sonication, the suspension was degassed under vacuum to remove entrapped air bubbles.

The membrane was fabricated via vacuum filtration using a Millipore assembly equipped with a PVDF membrane (0.4 µm pore size) as a supporting layer. The resulting wet cake was covered with a polyethylene sheet, sandwiched between filter papers, and cold-pressed to dry for two days to yield a freestanding membrane.

# NaVO synthesis

The NaVO material was synthesized in the same method as stated in our previous publication[4] and based on the technique of Wan et.al.[5]. In short, $NaV_3O_8 \cdot 1.5H_2O$ (NaVO) was synthesized by weighing 1 g $V_2O_5$ and mixing it with 15 mL 2 M NaCl and stirring for 96 h at 30°C until the solution turned dark red, indicating $Na^+$ insertion and nanorod formation[5,6]. The product was washed (water and ethanol, 5 times each) and dried at 60°C overnight.

**Cathode preparation**

NaVO, Super P, and PVDF (7:2:1 mass ratio) were mixed in NMP to form a slurry, cast on carbon paper (150 µm thickness), and dried at 60°C. Electrodes (⌀ 8 mm, loading 3.84 ±0.06 mg/cm²) were cut for testing.

# Characterizations

## Conductometric titrations

Conductometric titrations were carried out on the dialyzed fibers following a slightly adapted protocol to the one described by Foster et al.[7] And aliquot of the dialyzed sample containing 600 mg dry fibers was added to 500 mL degassed Milli-Q water. To better resolve the equivalence points, 2 mL of a 0.11 M ammonium acetate solution was added to the fibers, along with 7 mL of 0.1 M HCl solution to ensure the protonation of the ammonium and carboxylate groups. Thus-prepared samples were titrated **with 15** mL 0.1 M sodium hydroxide solution at 0.1 mL/min. For carboxymethylated deacetylated chitin fibers, the amount of HCl was increased to 15 mL, necessitating 25 mL of NaOH titrant.

Potentiometric datasets were obtained in the absence of ammonium chloride by monitoring the pH during a shorter conductometric titration experiment of 300 mg dry fibers acidified with 3 mL 0.1 M HCl and titrated with 7 mL 0.1 M NaOH at 0.1 mL/min.

To better compare the datasets obtained from conductometric titrations, we normalized the raw data by multiplying by the current volume of the analyte (the starting volume $V_o$ plus the volume of added titrant $V_T$) according to equation (1).

$$(S1)\ G_{norm} = (V_o + V_T) \cdot \sigma = (V_o + V_T) \cdot \sum_i c_i \Lambda_i = \sum_i n_i \Lambda_i \qquad \textbf{(1)}$$

$$Unit: [G_{norm}] = \frac{mS}{cm} \cdot L$$

$$(S2)\ \sigma_{1L} = \sum_i \frac{(V_o + V_T)}{1L} c_i \Lambda_i = \frac{\sum_i n_i \Lambda_i}{1L} \quad \textbf{(2)}$$

The calculated conductance $G_{norm}$ has the same numeric value as the conductivity of a solution containing the same amounts of analyte electrolytes in a standard volume of 1 L, as expressed in equation (2). Additionally, the normalized conductance reflects changes in the amounts of the contained ionic species i, $n_i$, while the measured conductivities reflect the respective concentrations $c_i$. Therefore, the normalized conductance curves are directly comparable, given that the analyzed samples contain similar amounts if not concentrations of analytes and auxiliaries.

## Atomic force microscopy

The nanochitin suspensions were diluted to 0.005 wt% using water at pH 3. Subsequently the samples were spin coated (4000 rpm for 1minute) on a cleaned silicon wafer and imaged using a Bruker Multimode 8 AFM. The images were processed using Gwyddion and height and length distributions were extracted using Fibre App[8].

## ζ-potential analysis

The ζ-potential of the different chitin fibril suspensions was measured as a function of pH using Zetasizer Nano Zs (Malvern Instruments, U.K.). A concentration of 0.1 wt.% of nanochitin suspension was used, and pH was adjusted with 0.1M HCl and NaOH. Three measurements were performed to obtain the electrophoretic mobility of the particles. The ζ-potential was then calculated using Smoluchowski theory, which is valid for spherical particles. Consequently, the obtained values should be considered representative, rather than absolute.

## Mechanical properties of chitin separators

The mechanical properties of the chitin separators were evaluated using an Instron 4204 Universal Tester (USA) equipped with a 1 kN load cell. To prepare for testing, the separator membranes were punched into a dog-bone geometry and equilibrated in a 1M Zn(OTf)2 solution for 24 hours. Tensile testing was conducted in accordance with ASTM D882 guidelines for thin films to determine the specimens' tensile strength and modulus.

## Scanning electron microscopy

The recovered zinc electrodes and separators were analysed using a LEO 1550 Gemini Zeiss SEM at 5 kV. Prior to imaging, the samples were vacuum-dried for 10 minutes and the separators were sputter-coated with gold. For the SEM comparison, new cells were run at 1 mA/cm$^2$ and 1 mAh/cm$^2$ for 20 cycles ending with a discharge before disassembly.

## Raman spectroscopy

Raman spectra were recorded on a WITec Alpha 300M+ system in confocal backscattering mode using a 20× objective. Measurements employed a 532 nm laser (50 mW power), with 0.5 s integration time, 60 accumulations, and 2 cm$^{-1}$ resolution. The electrolyte was held in a 2 mm quartz cuvette during analysis.

### Electrochemical impedance spectroscopy

The frequency sweep was from 100 kHz to 50 mHz with a potential amplitude of 10 mV for all experiments. The ionic resistances were determined as the curve intercepts with the real part of the impedance axis.

### Ionic conductivity

The ionic conductivity was determined through potentio electrochemical impedance spectroscopy (PEIS) measurements. For the electrolyte, a two-electrode cell was calibrated, and the cell constant $k_C$ was determined accordingly. The $k_C$ was in the range of 0.9 – 1.1. The ionic conductivity of the separators was determined by utilising the BioLogic CESH-e sample holder in the through-plane configuration. The measurements were conducted in a BioLogic ITS-e temperature control unit in the temperature range of 20 to 40 $^{o}$C, with a stabilisation time of 1 hour at each temperature.

### X-ray diffraction

X-ray diffraction (XRD) analysis was conducted on a Rigaku SmartLab diffractometer with Cu Kα radiation (λ = 1.5406 Å). The samples were scanned at a 2°/min rate with a step size of 0.05°.

### Neutron scattering

The QENS measurements were carried out at the Institute Laue-Langevin (Grenoble, France) by using the direct geometry disc chopper cold neutron time-of-flight spectrometer IN5, operating at a wavelength of 5 Å. This setting gives a resolution at the elastic line of ~ 0.1 meV and a Q range of ~ 0.2-2.3 $Å^{-1}$. Samples were sealed in annular cylindrical containers with an optimized thickness of 0.5 mm, relevant to the minimization of effects like multiple scattering and absorption. The Mantid software[9] was used to reduce and treat the data, and subsequently to extract the full Q-averaged QENS spectra, S(E).

### X-ray photoelectron spectroscopy

X-ray photoelectron spectroscopy (XPS) analyses were carried out using a Thermo Fisher Scientific K-Alpha spectrometer, and the resulting data were processed with Avantage software. Zinc foils were mounted on carbon tape affixed to the sample holder and analyzed using monochromatic Al Kα radiation (1486.6 eV). Survey spectra were collected with a pass energy of 200 eV and a step size of 1 eV, using an analysis spot of approximately 100 µm. High-resolution spectra were acquired with a pass energy of 50 eV and a step size of 0.1 eV. Binding energies were referenced to the adventitious carbon C-C peak at 284.8 eV[7]. A U2 Tougaard background was applied, and peak fitting was performed using a mixed Gaussian-Lorentzian line shape (GL(30)). The electron kinetic energies across the analyzed regions ranged from approximately 480 to 1280 eV, corresponding to electron inelastic mean free paths (IMFPs) of around 0.7-1.9 nm and an effective probing depth of around 2.1-5.7 nm (3 × IMFP)[8].

## Electrochemical measurements

### 2D/3D plating

The 2D/3D Zn plating behaviour was evaluated using a two-compartment Swagelok cell with Ø 10 mm Zn foil as both the working (WE) and counter electrodes (CE), a Ø 13 mm separator, and 25–40 µL of electrolyte for chitin-based separators or 90 µL for glass fiber and cellulose separators. After 10 hours of stabilization, chronoamperometry was performed by applying -150 mV for 10 minutes.

## Coulombic efficiency

A Swagelok cell with Ø 12 mm Cu (WE), Ø 10 mm Zn (CE) and Ø 13 mm separator was used. The electrolyte volumes were 25-40 uL and 90 uL for the chitin-based and glass fiber or cellulose, respectively. The current density and capacity were 1 mA/cm$^2$ and 1 mAh/cm$^2$ throughout the whole experiment.

## Cyclic voltammetry

Cyclic voltammetry was conducted in a two-compartment Swagelok cell with a Ø 12 mm Cu (WE), Ø 10 mm Zn (CE), and Ø 13 mm separator. The applied current density was 1 mV/s.

## Desolvation energy

Desolvation energy ($E_{des}$) was measured using two-compartment Swagelok cells with Ø 10 mm Zn electrodes (WE and CE), a Ø 13 mm separator, and 25–40 µL (chitin-based) or 90 µL (glass fiber/cellulose) electrolyte across 20–60 °C. After a 10-hour rest and 5 cycles at 1 mA/cm$^2$ and 1 mAh/cm$^2$, the cell was equilibrated at each temperature for 1 hour before PEIS was conducted. The acquired data were fitted with the Arrhenius equation (S3) to obtain the desolvation energy $E_{des}$.

$$(S3)\ Ln(R_{ct}^{-1}) = \exp\left(\frac{-E_{des}}{RT}\right)$$

## Zn||Zn stability

Cycling stability of Zn‖Zn cells was evaluated using two-compartment Swagelok cells with Ø 10 mm Zn electrodes (WE and CE), a Ø 13 mm separator, and 25–40 µL (chitin-based) or 90 µL (glass fiber/cellulose) electrolyte. After a 10-hour rest, long-term tests were performed under varying conditions: 0.5 mA/cm$^2$ with 0.4 mAh/cm$^2$, 1 mA/cm$^2$ with 1 mAh/cm$^2$, 1 mA/cm$^2$ with 5 mAh/cm$^2$, and 5 mA/cm$^2$ with 5 mAh/cm$^2$. The cutoff potential was set to 0.5 V.

## Zn||NaVO full cell stability

Full cell experiments with Zn‖NaVO were performed in a two-electrode Swagelok cell using NaVO with Ø 8 mm as WE, Zn foil with Ø 10 mm as CE, and a Ø 13 mm separator with 25-40 µl of electrolyte for the chitin-based separators or 90 µl for the glass fibre and cellulose separators. The rate capability test was conducted by applying varying current rates from 0.1 A/g to 5 A/g, and finally returning to 0.1 A/g. The long-term cyclability was performed under an applied current rate of 2 A/g.

# Figures S1-23

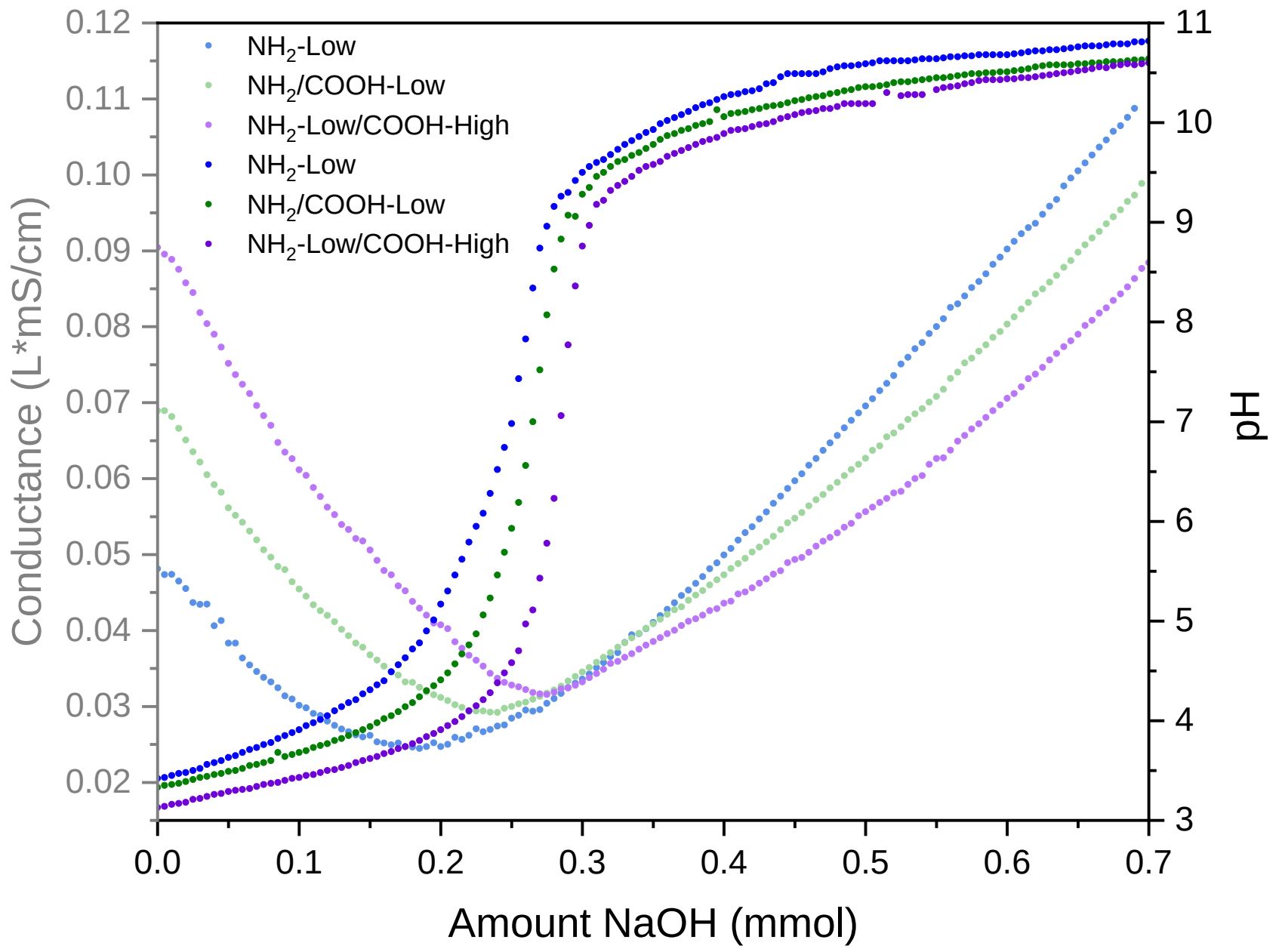

*Fig. S1: Potentiometric and conductometric titration curves of deacetylated and carboxymethylated fibers. The $NH_2$-Low/COOH-High fibers have not been used for the membranes; they are only used to establish the synthesis and titration method.*

**Carboxymethylation confirmed by conductometric titration**

Neither the potentiometric nor the conductometric titrations of the neat, washed, and dialyzed fibers (Fig. S1) reveal any unambiguous equivalence points to quantify any surface groups. Instead, the curves as a whole are shifted following the trend in their surface charge. In their protonated state, the deacetylated fibers have the largest surface charge and the largest acidity. The subsequent introduction of oppositely charged carboxylate moieties reduces the surface charge and consequently reduces the surface acidity. This is reflected in the initial pH and conductance values. The more basic fibers bind more protons, raising the pH of the analyte solution and decreasing its conductivity. Accordingly, the pH in the system increases faster with basic fibers. This, too, is a consequence of the surface charge that has been studied for stiff polycarboxylates[2] and nanoparticles carrying carboxylate groups[3]. The surface potential influences the conventional dissociation equilibrium and modulates the surface groups' dissociation. As such, a positive surface charge increases the surface acidity, as the dissociation of a proton decreases the surface charge, either by converting an ammonium group into a neutral amine or by converting a neutral carboxylic acid moiety into a negatively charged carboxylate[2,3]. As such, the titration curves confirm a reduction in surface potential ensuing carboxylation.

To assess the changes on a more quantitative basis, we introduced a known amount of ammonium acetate. As we are dealing with a mixture of amines and carboxylates on the fiber surfaces, acetate and ammonium ions buffer the pH increase in a very similar range expected for surface carboxylates and amines in the absence of surface potential, respectively. Therefore, the addition of acetate and ammonium ions artificially introduces easily detected equivalence points in the recorded conductometric titration curves, as shown in Fig. S2.

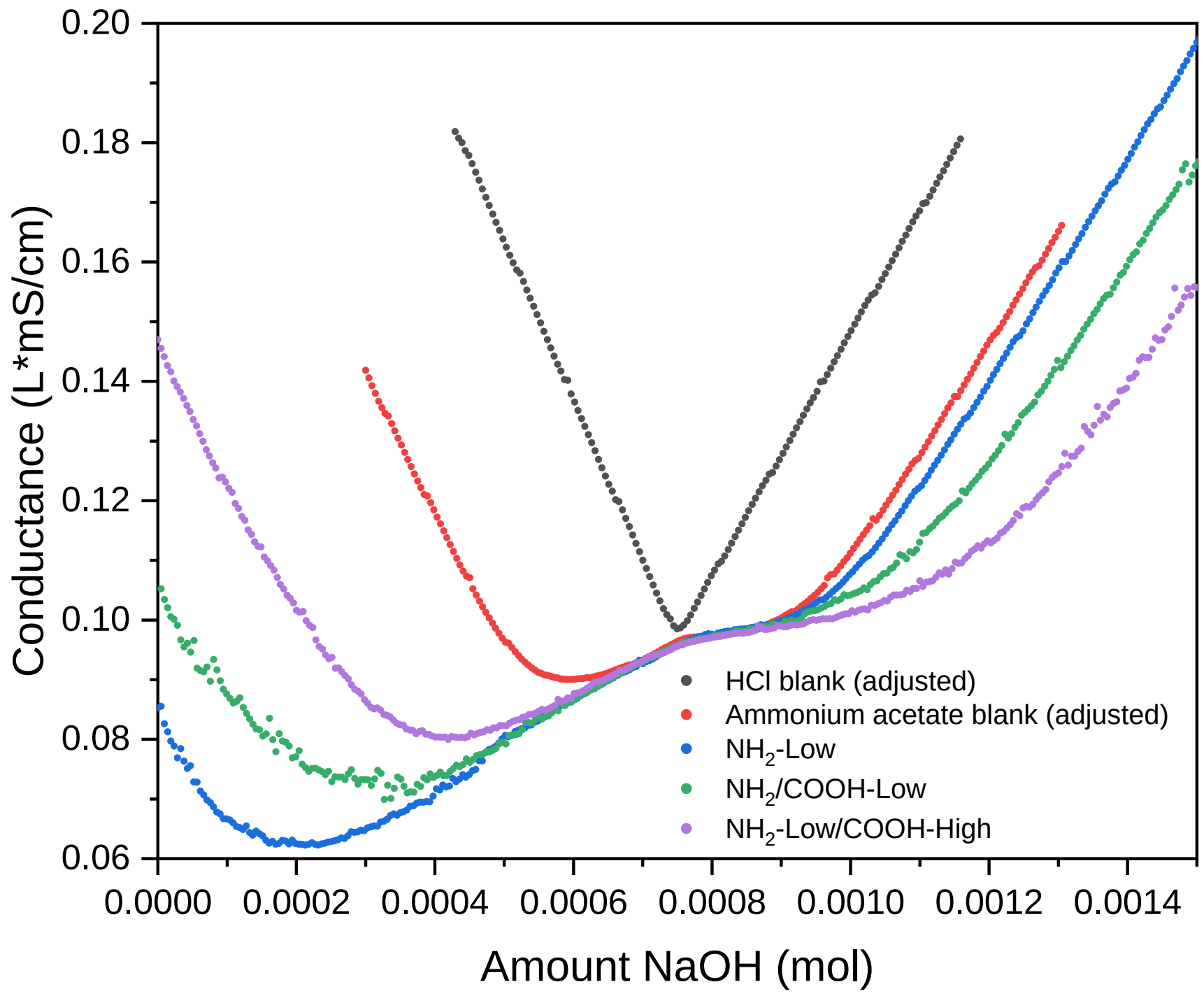

*Fig. S2: Normalized conductance curves of low-charged deacetylated and carboxymethylated fibers ($NH_2$-Low and $NH_2$/COOH-Low) in the presence of ammonium acetate. The curves containing no fibers were adjusted by considering discrepancies in the amounts of acid (shift on the x-Axis by equivalent amounts NaOH) and ionic strength, which induces higher conductivity (shift on the y-Axis by 114 (mS/cm)/(mol/L) NaCl and 95 (mS/cm)/(mol/L) $NH_4OAc$ at 20 °C and neutral pH). The $NH_2$-Low/COOH-High fibers have not been used for the membranes; they are only used to establish the synthesis and titration method.*

The presence of ammonium acetate increases the overall conductivity of the system and broadens the titration curves by increasing the sodium hydroxide consumption. One equivalent of ammonium acetate binds one equivalent of protons when acidified with HCl and consumes one additional equivalent of sodium hydroxide upon neutralization. This is illustrated by comparing the blank titration curves of HCl, displaying the characteristic V-shape observed for strong acids, and that of ammonium acetate acidified with HCl. Our samples contain 0.22 mmol ammonium acetate, which accounts for the consumption of 0.44 mmol NaOH. The remaining width of the curves between their two induced equivalence points is due to deacetylation and carboxymethylation.

Additionally, due to the normalization of the datasets, the slopes of the curves match during the initial neutralization of free protons and the addition of excess NaOH towards the end. This consistency demonstrates the comparability of the individual curves. Furthermore, given that the fibers have been treated identically during washing and contain the same amounts of HCl and ammonium acetate, their normalized titration curves match each other exactly during the transition from weakly acidic to weakly basic conditions.

The differences in the titration curves of the fibers highlight the effect of the surface deacetylation and carboxymethylation. Deacetylation exposes basic amine functionalities on the fibers, which are protonated in acidic conditions. Therefore, the initial conductance of the analyte is reduced. This leads to a positive surface charge, which, in turn, increases the acidity of the (protonated) ammonium moieties. This effect is reflected in the minimal widening of the titration curve in basic conditions compared to the large capacity for proton capture in acidic conditions. The introduction of carboxylate moieties on the surface mitigates this effect. The protonation of the basic amine functionalities and subsequent surface charge increases the acidity of both the ammonium groups as well as the carboxylates. At sufficiently high charge densities, therefore, carboxylates may behave like strong acids and dissociate freely. Dissociated, negatively charged carboxylates, in turn, compensate the positive surface charge and thereby lower the acidity of the protonated groups on the fiber surface. Consequently, with increasing amounts of carboxymethylation, the fibers bind fewer protons due to increased overall acidity of the material. Simultaneously, full deprotonation requires more sodium hydroxide, due to decreased acidity of the ammonium groups, particularly in the presence of excess negative surface charge.

Separately from these charge-induced influences, the titration curves are shifted inherently due to autoprotolysis of the fibers. The coexistence of carboxylic acid moieties next to amine groups leads to the formation of ammonium carboxylates via protolysis, which means that compared to the purely deacetylated fibers, the carboxymethylated fibers inherently retain protons. Accordingly, the pH jumps in the potentiometric titration curves in Figure S1 are shifted towards larger amounts of NaOH.

The variable acidic strength of the fibers makes it difficult to reliably quantify the total functional group content and, more difficult still, to differentiate between carboxylate and amine moieties. Still, we can quantify the weakly acidic functional groups. We will interpret the shift in the titration curves as a direct result of the presence of carboxylate moieties. Therefore, the total functional group content is determined as the sum of this carboxylate shift and the weakly acidic groups. The results are listed in Table S1.

On the whole, the amount of weakly acidic groups stays the same in the samples, which is what we expect for carboxymethylation: The amount of amine groups stays the same, they are just transformed into secondary amines. The formation of tertiary and quaternary amines is not expected due to steric inhibition of successively more substituted amines. Therefore, the basic sites are expected to be retained after modification. The carboxylate groups, in turn, increase the degree of initial protonation and appear to dissociate significantly above pH 3. The opposite effect of the negative charge of the introduced carboxylate groups on the dissociation of the protonated amines at high pH is evident at high carboxylate content from the significantly reduced slope of the titration curve. Therefore, the total amount of weakly acidic sites may be underestimated.

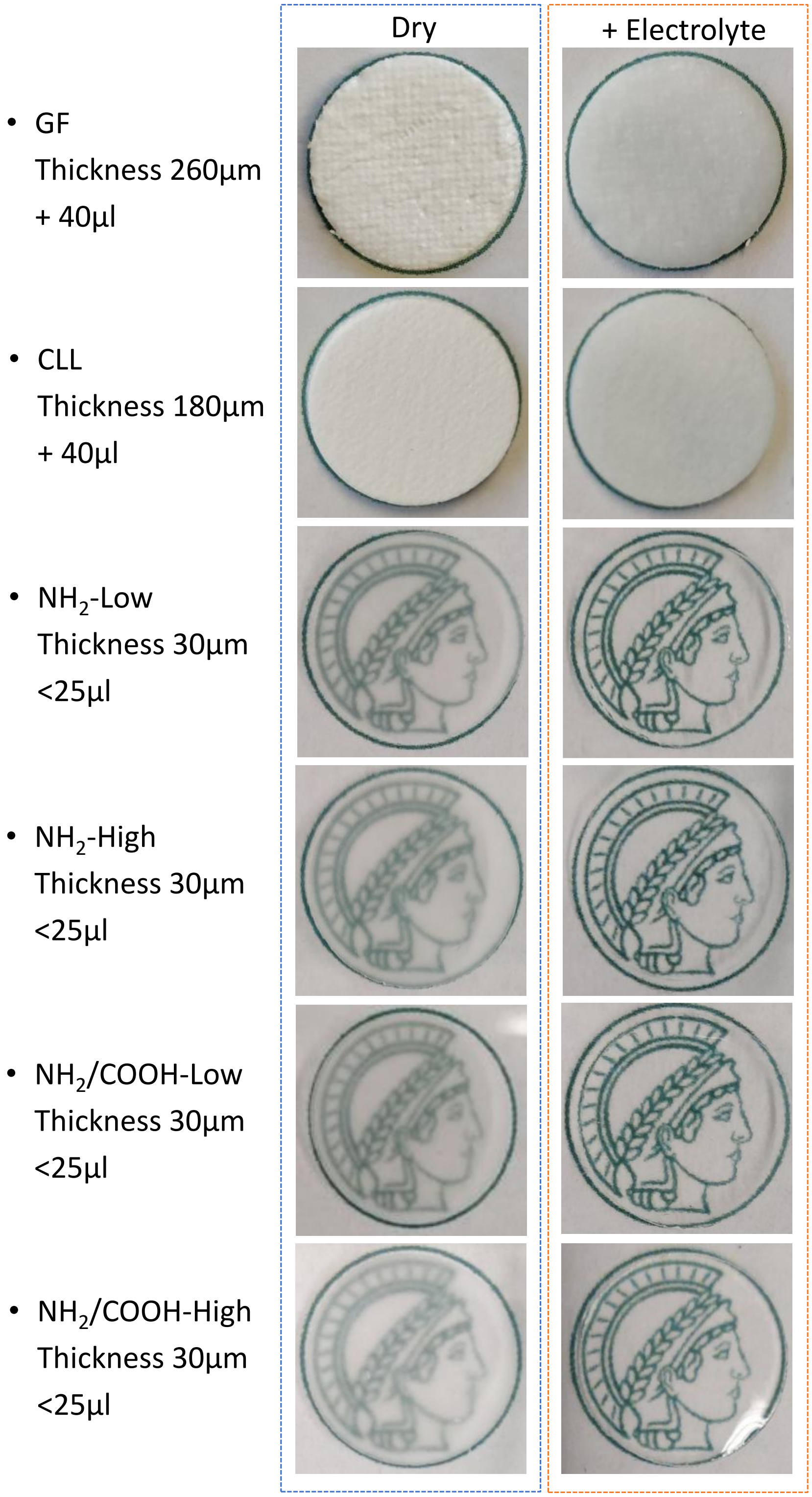

*Fig. S3: Wettability of the separators.*

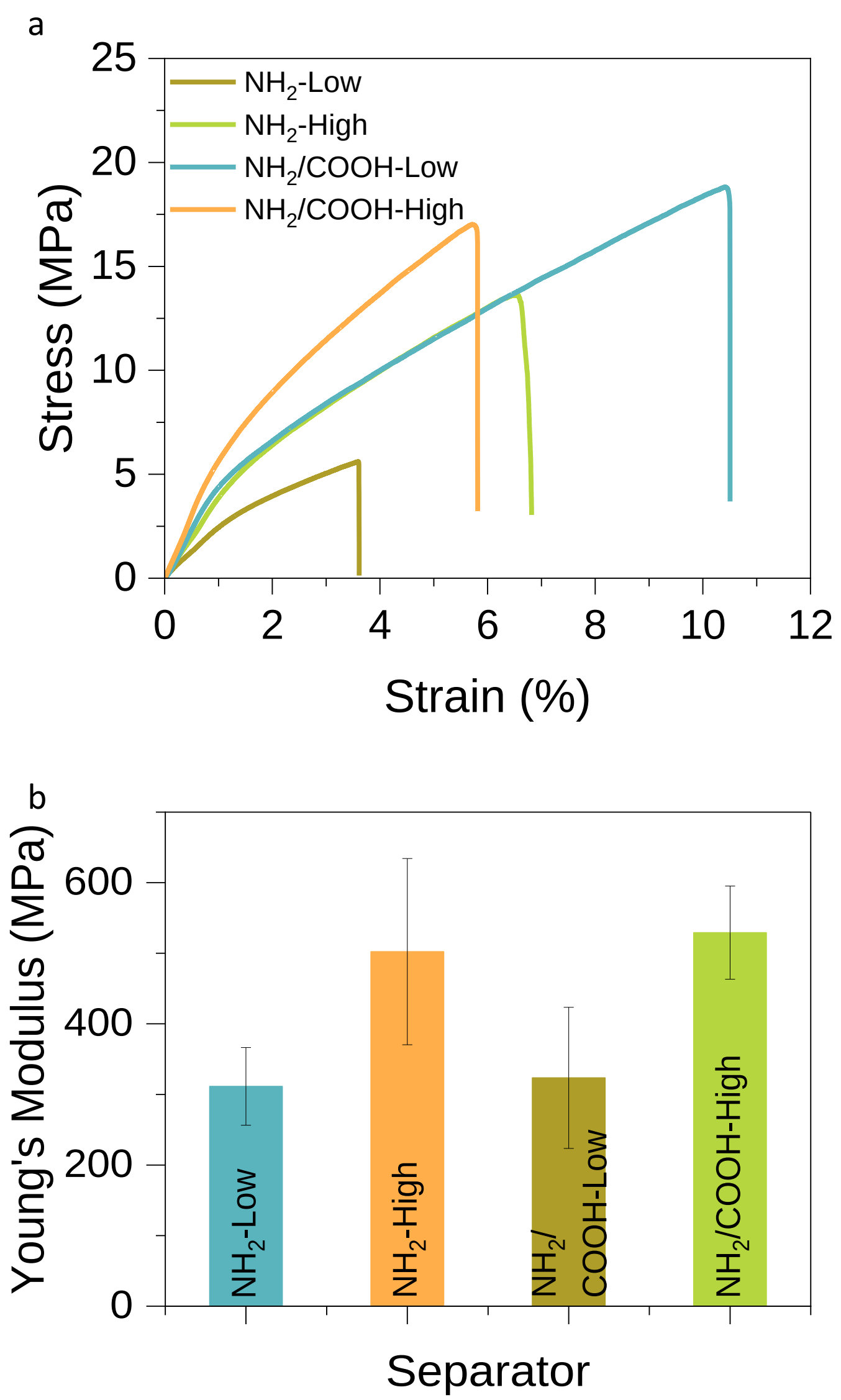

*Fig. S4: Mechanical properties of the proposed separators. a) Stress-strain curves and b) Young's modulus of separator membranes equilibrated in 1M Zn $(OTf)_2$ electrolyte.*

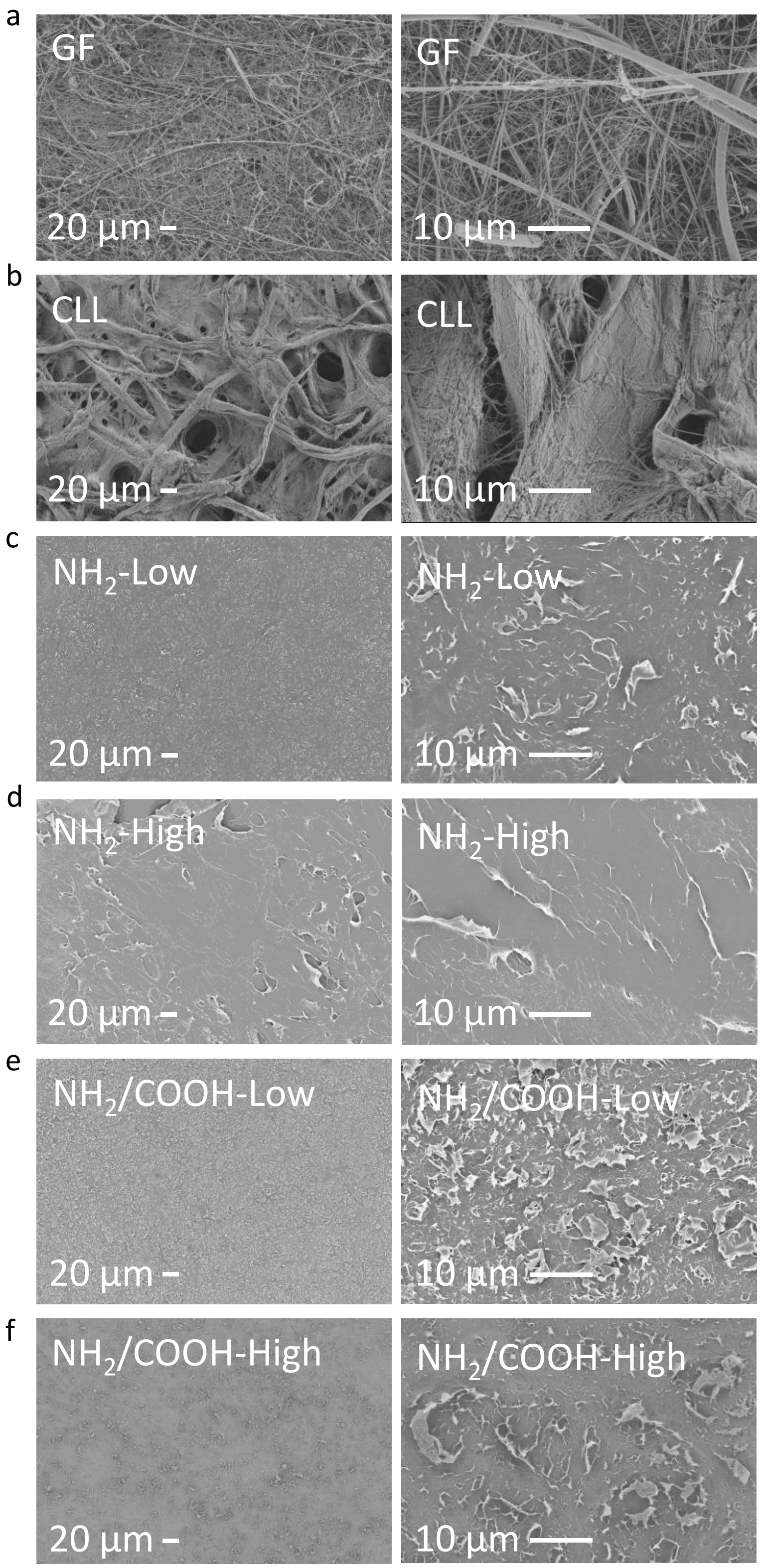

*Fig. S5: SEM images of the pristine separators at different magnifications.*

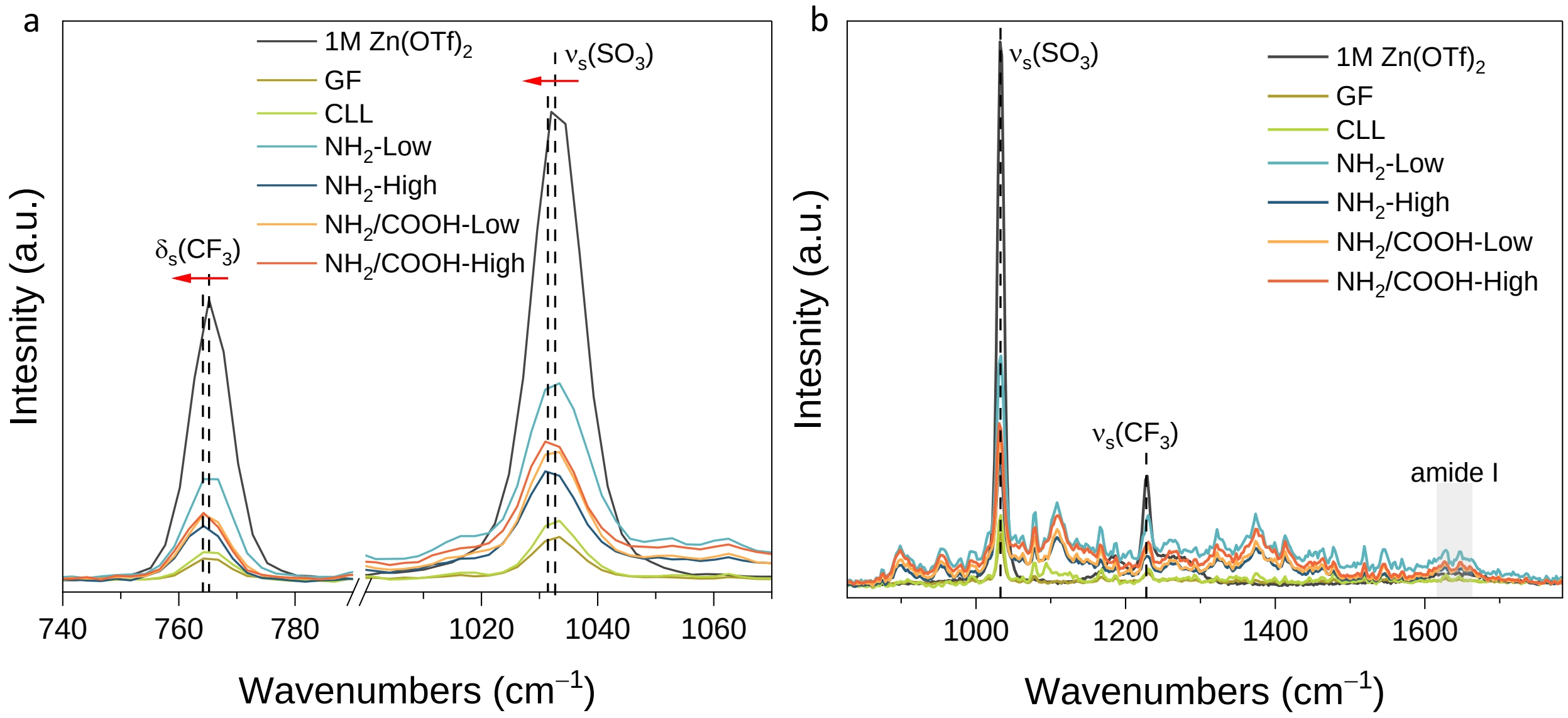

*Fig. S6: Raman analysis of the separators in the Raman range from a) 740-1070 cm$^{-1}$ and b) 850-1750 cm$^{-1}$. ν = bond stretch, δ = scissoring (in-plane-bending), subscript s = symmetric.*

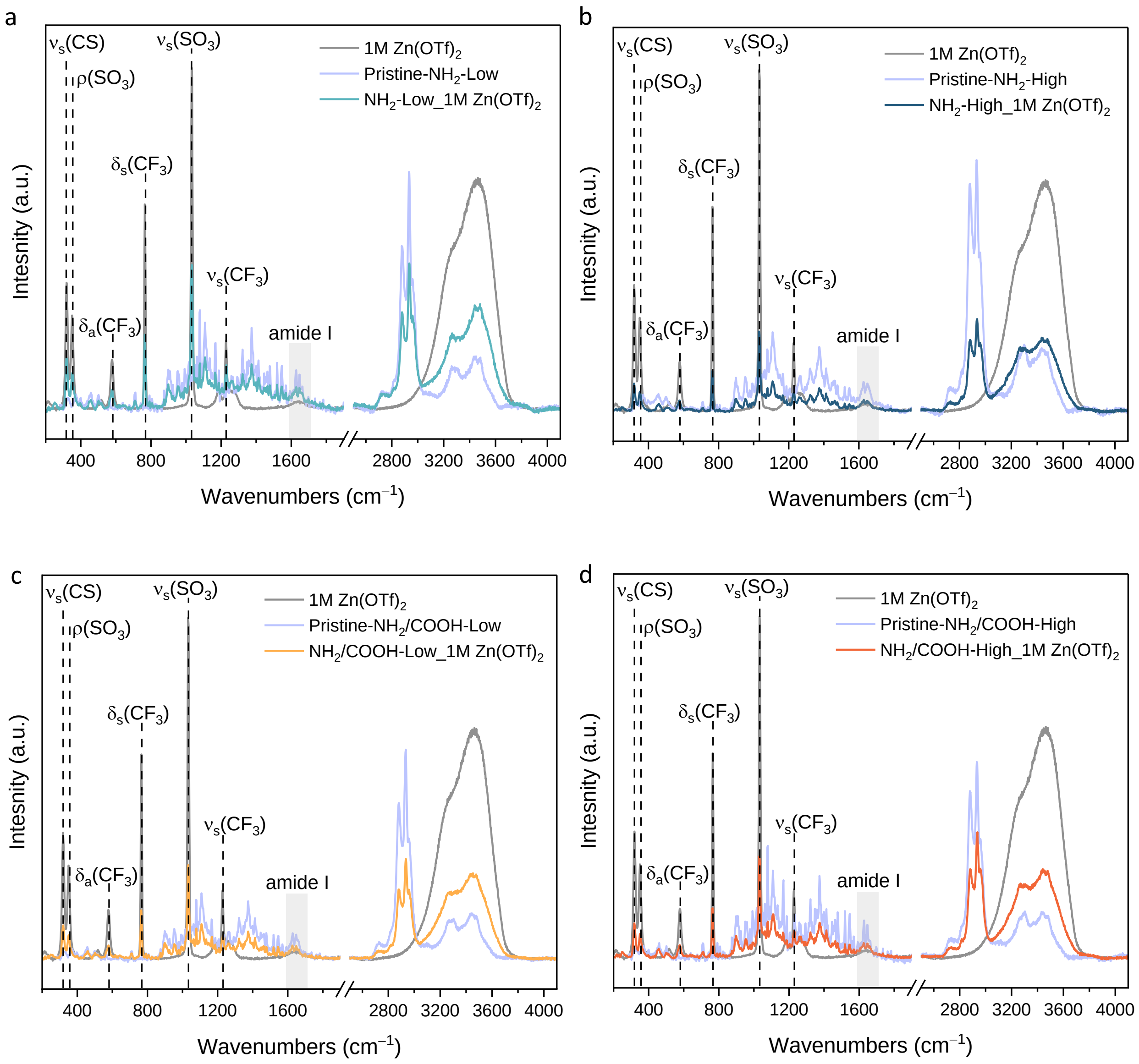

*Fig. S7: Full Raman spectra of the pristine and activated (addition of 1 M Zn(OTf)$_2$) chitin nanofibre separators compared to 1 M Zn(OTf)$_2$ electrolyte. a) NH$_2$-Low, b) NH$_2$-High, c) NH$_2$/COOH-Low, and d) NH$_2$/COOH-High. ν = bond stretch, δ = scissoring (in-plane-bending), ρ = rocking, subscript s = symmetric.*

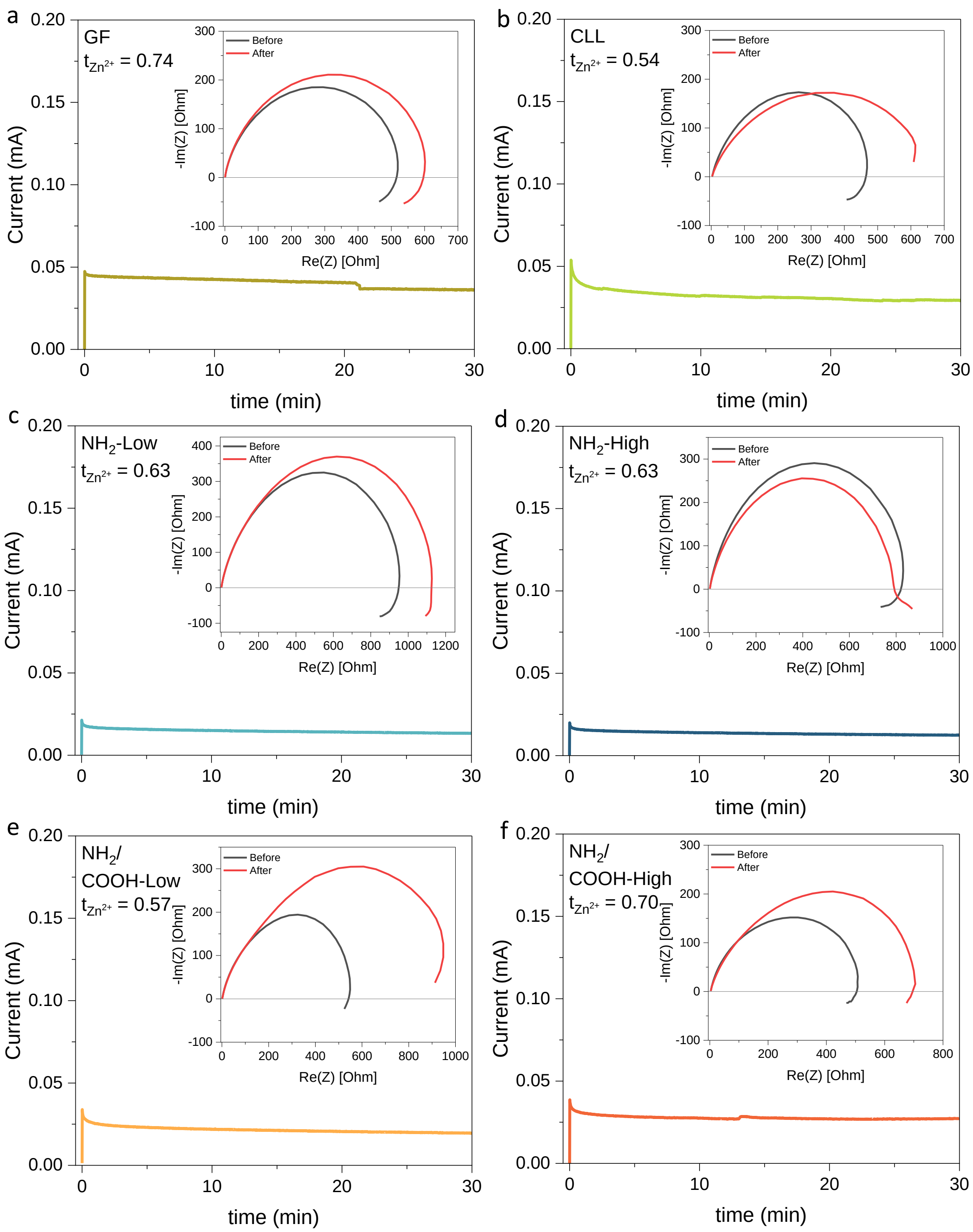

*Fig. S8: Chronoamperometry and EIS spectra for determining the transference number at 30 mV. a) GF, b) CLL, c) $NH_2$-Low, d) $NH_2$-High, e) $NH_2$/COOH-Low, and f) $NH_2$/COOH-High. g) comparison of the determined $t_{Zn2+}$.*

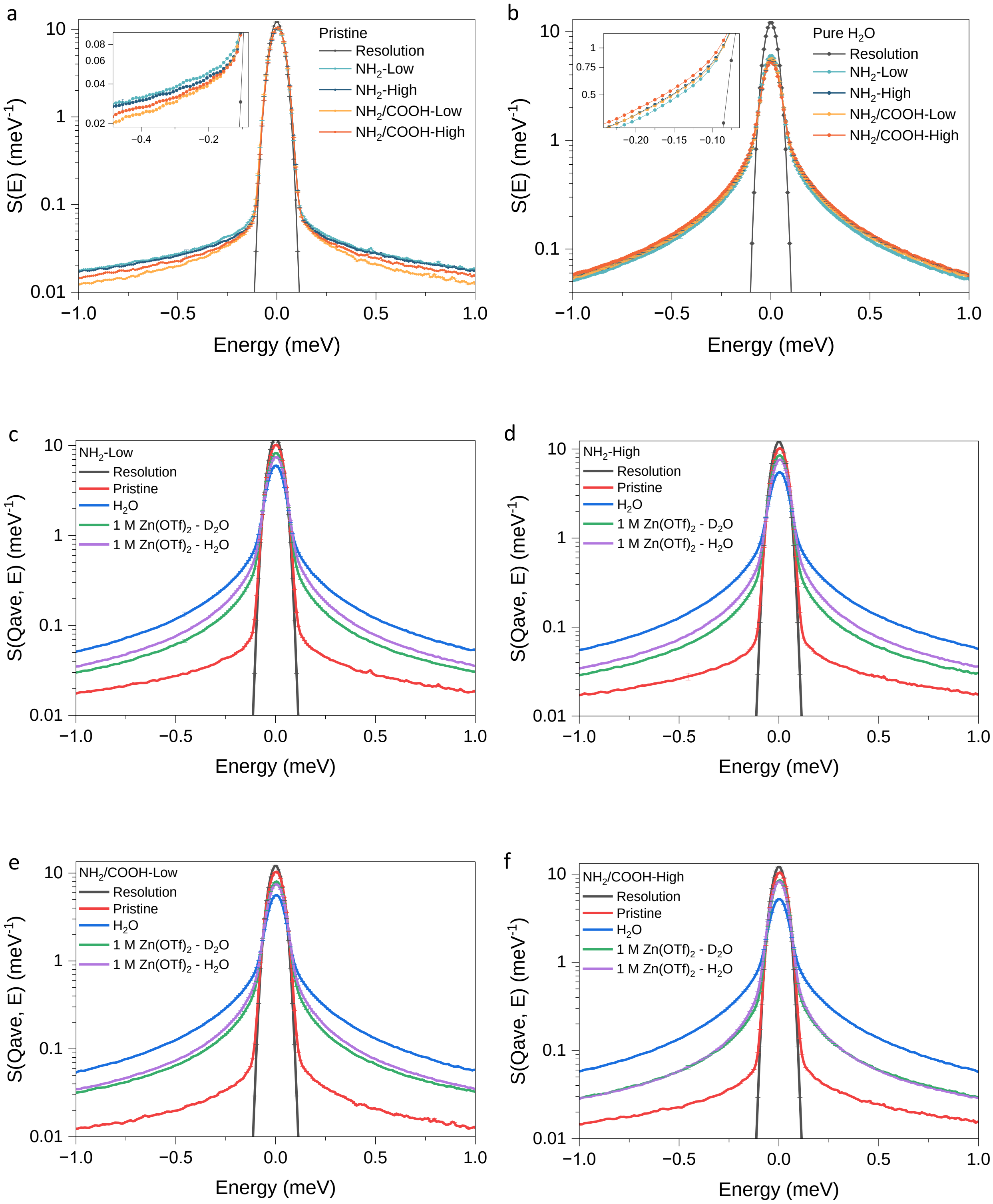

*Fig. S9: QENS spectra with $Q_{ave}$ of a) pristine separators, b) separators in only water, c) $NH_2$-Low, d) $NH_2$-High, e) $NH_2$/COOH-Low, and f) $NH_2$/COOH-High.*

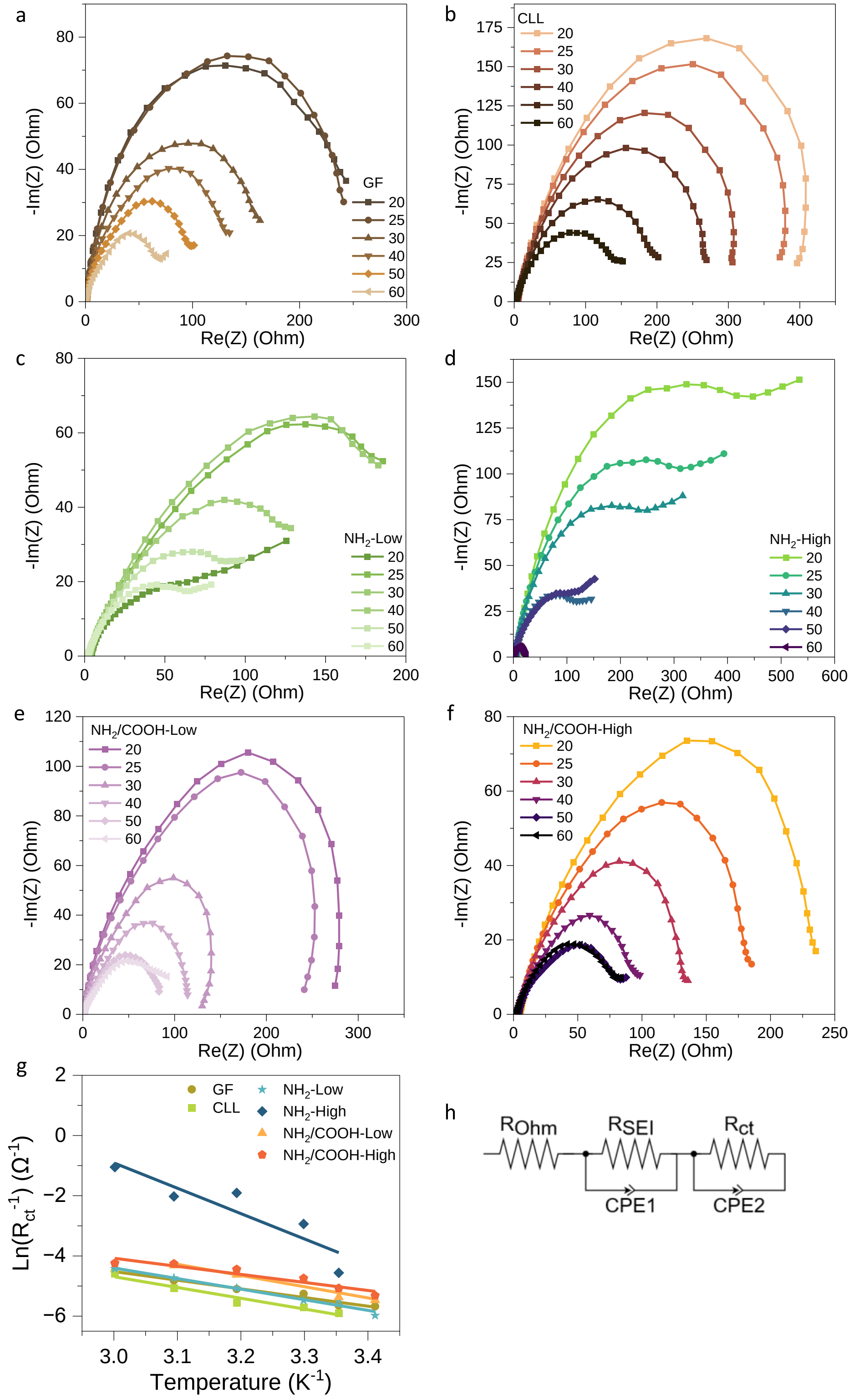

*Fig. S10: Determination of the desolvation energies through EIS measurements. a-f) Nyquist plots of Zn symmetrical Zn||Zn cell with the respective separator under different temperatures. g) Arrhenius curves. h) The fitted circuit to the Nyquist plots.*

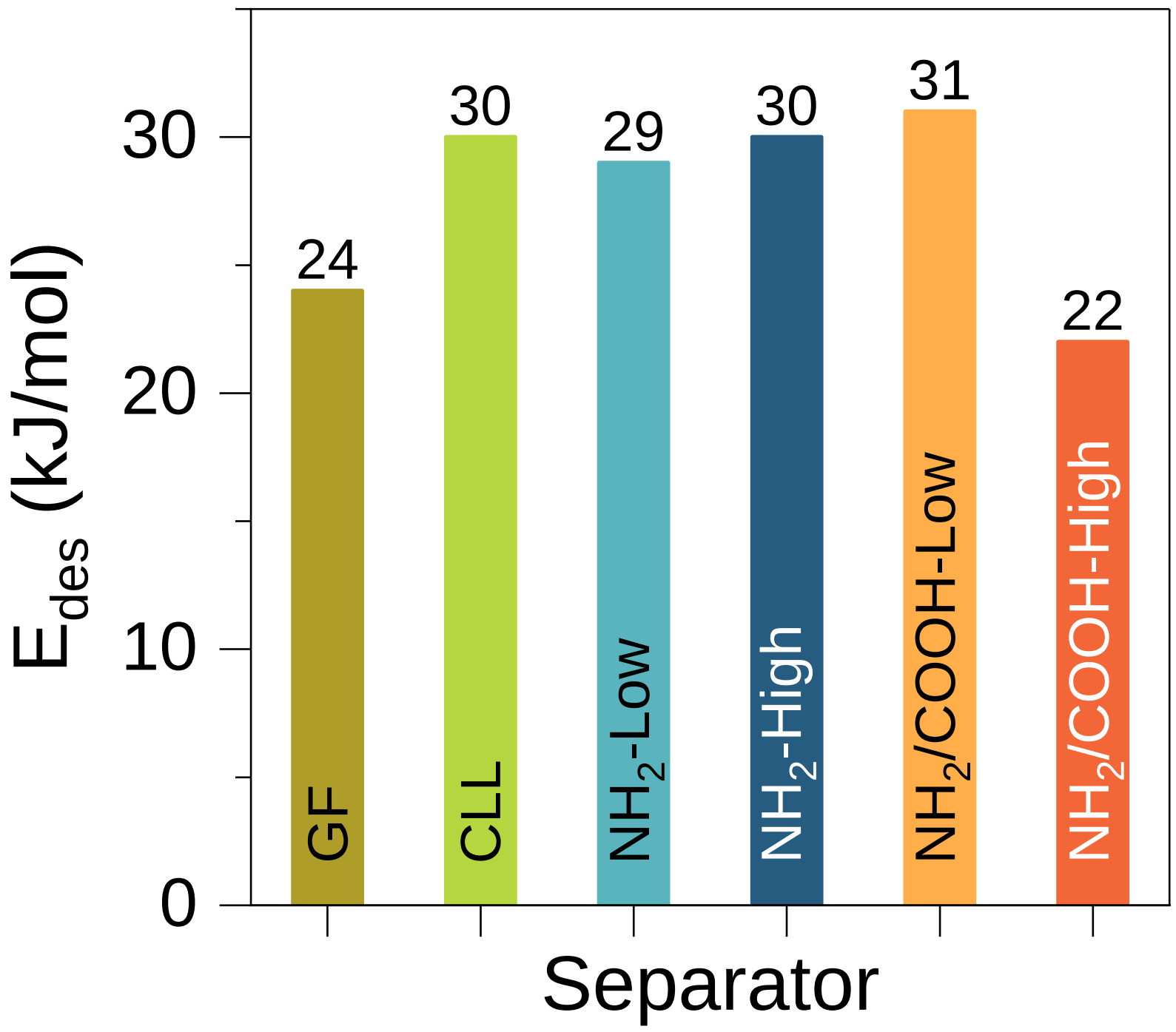

*Fig. S11: Fig. S11: Determined desolvation energies.*

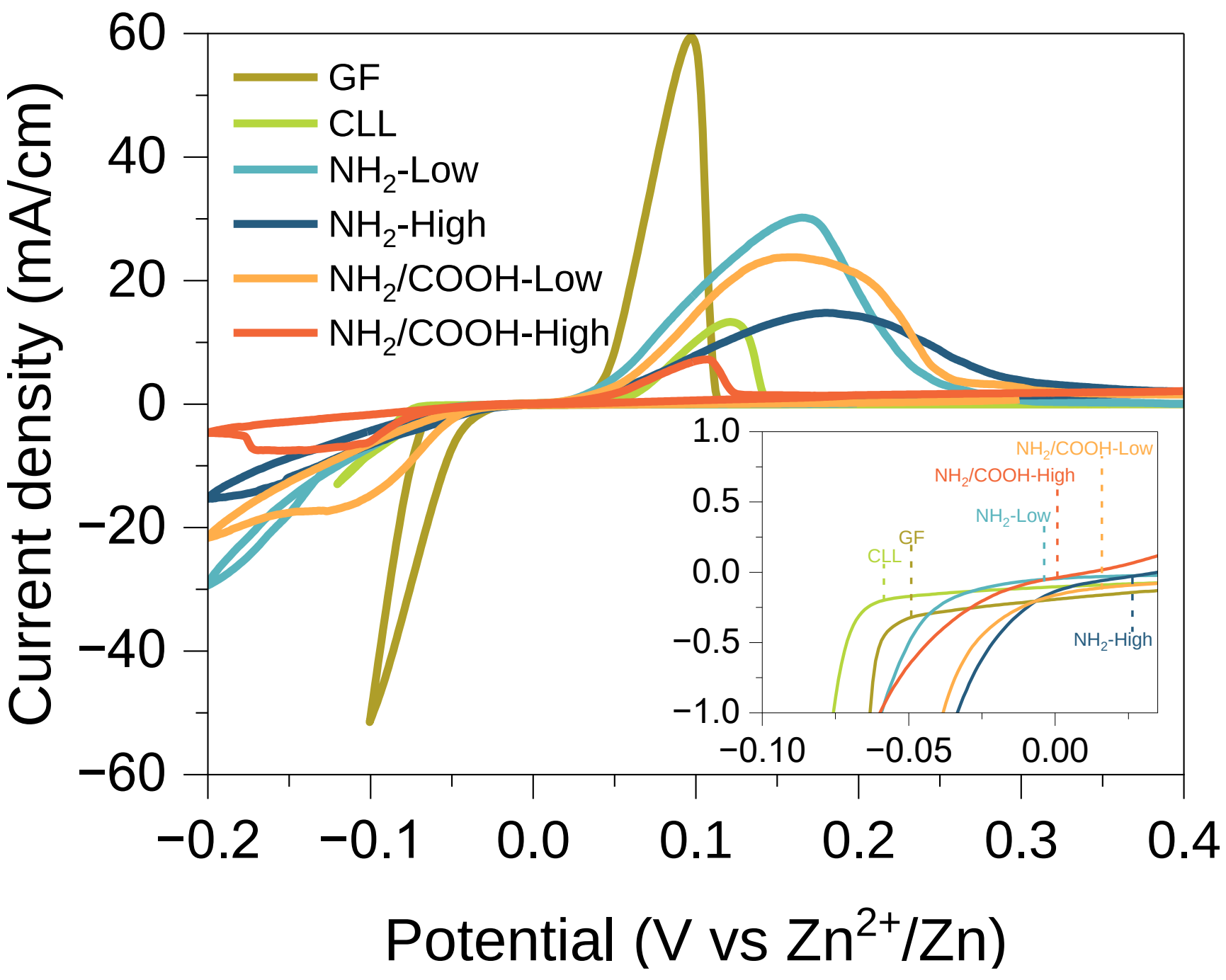

*Fig. S12: Acquired cycling voltammetry curves in a Zn‖Cu cell with a current density of 1 mV/s.*

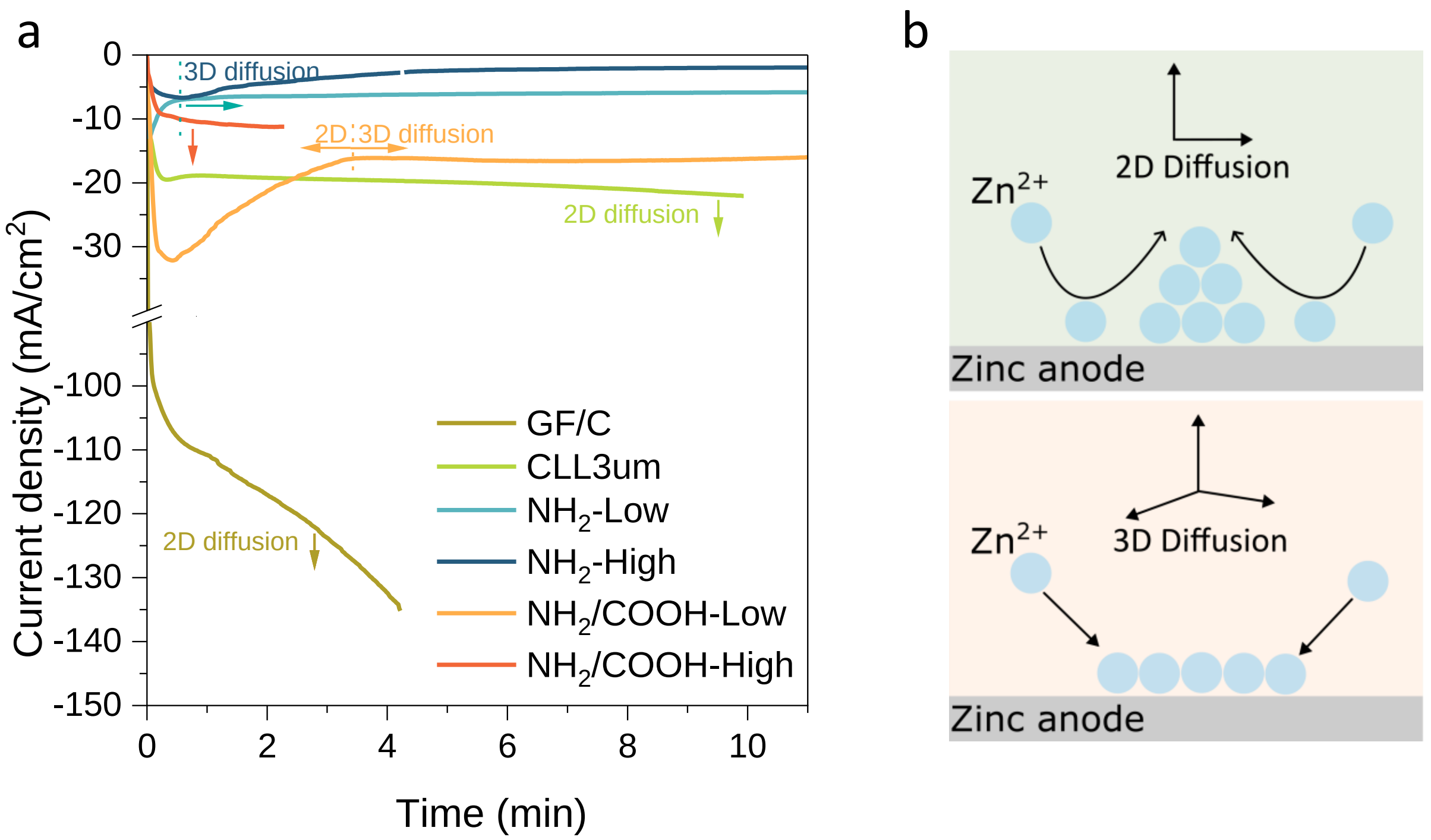

*Fig. S13: Determining the 2D/3D deposition behavior. a) CA curves at a potential of -150 mV and b) schematic drawing of 2D/3D deposition.*

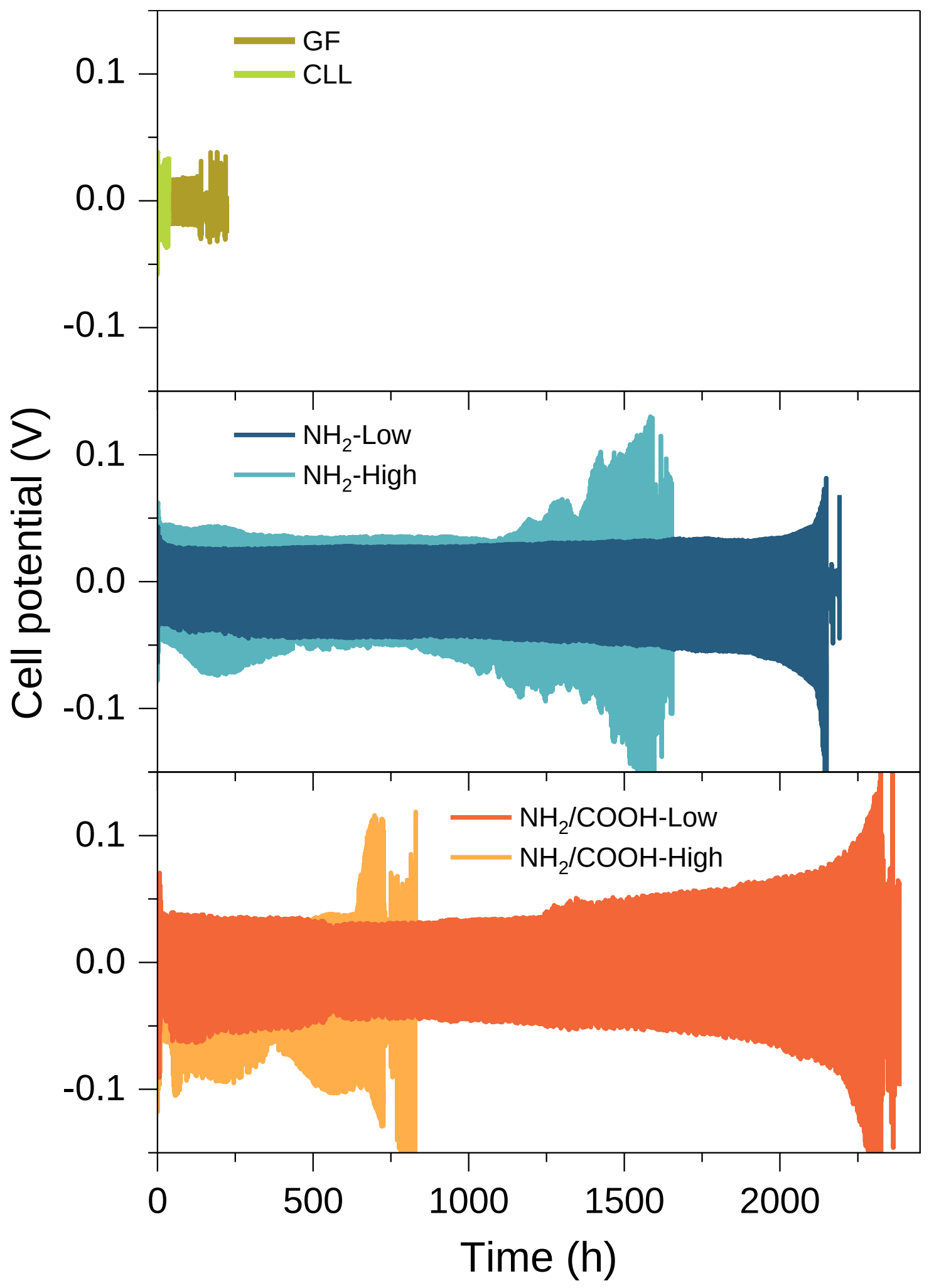

*Fig. S14: Long-term cycling of the Zn anode with 1 M Zn(OTf)$_2$ at 0.5 mA/cm$^2$ and 0.4 mAh/cm$^2$.*

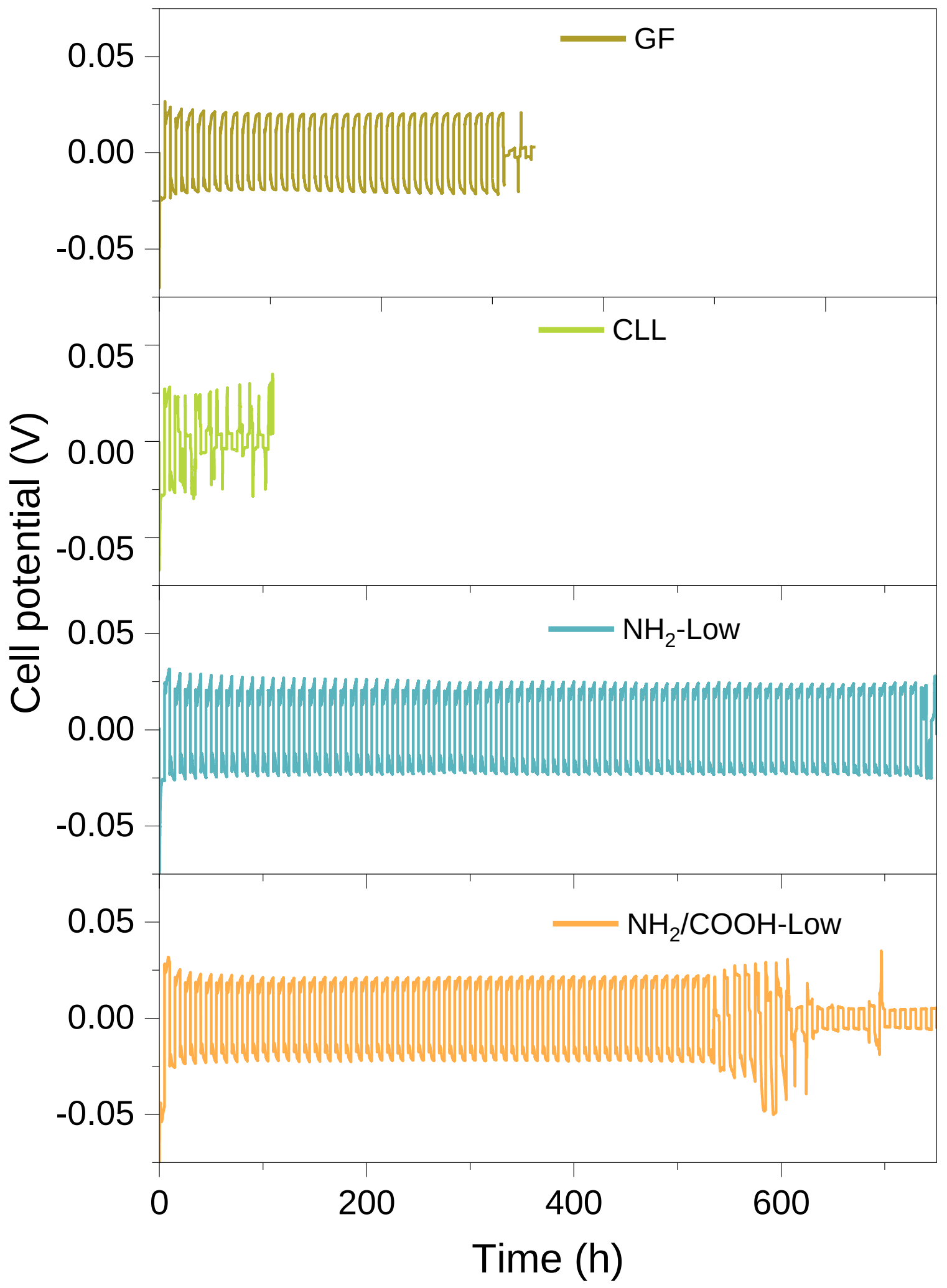

*Fig. S15: Long-term cycling of the Zn anode at 1 mA/cm$^2$ and 5 mAh/cm$^2$.*

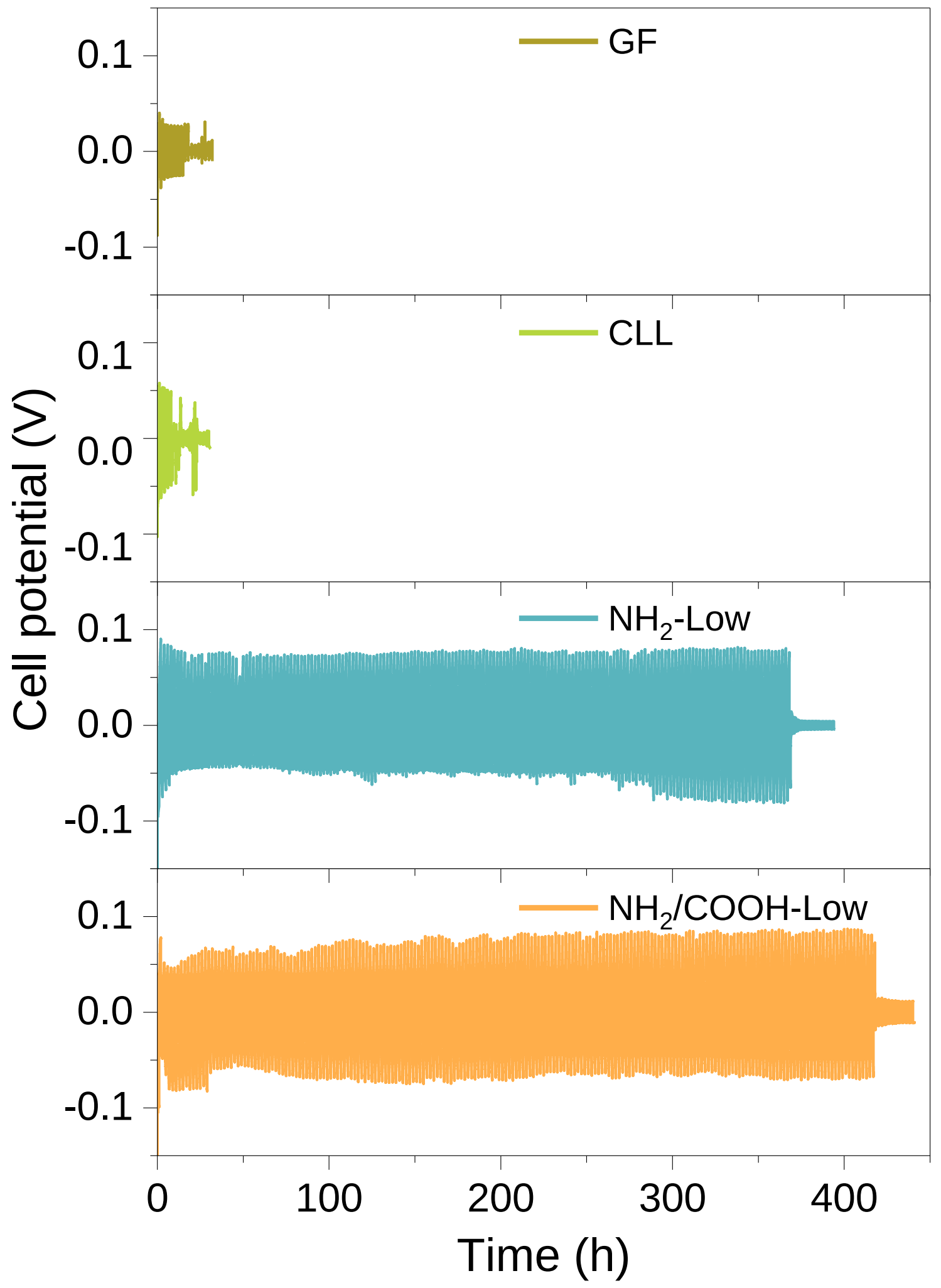

*Fig. S16: Long-term cycling of the Zn anode at 5 mA/cm$^2$ and 5 mAh/cm$^2$.*

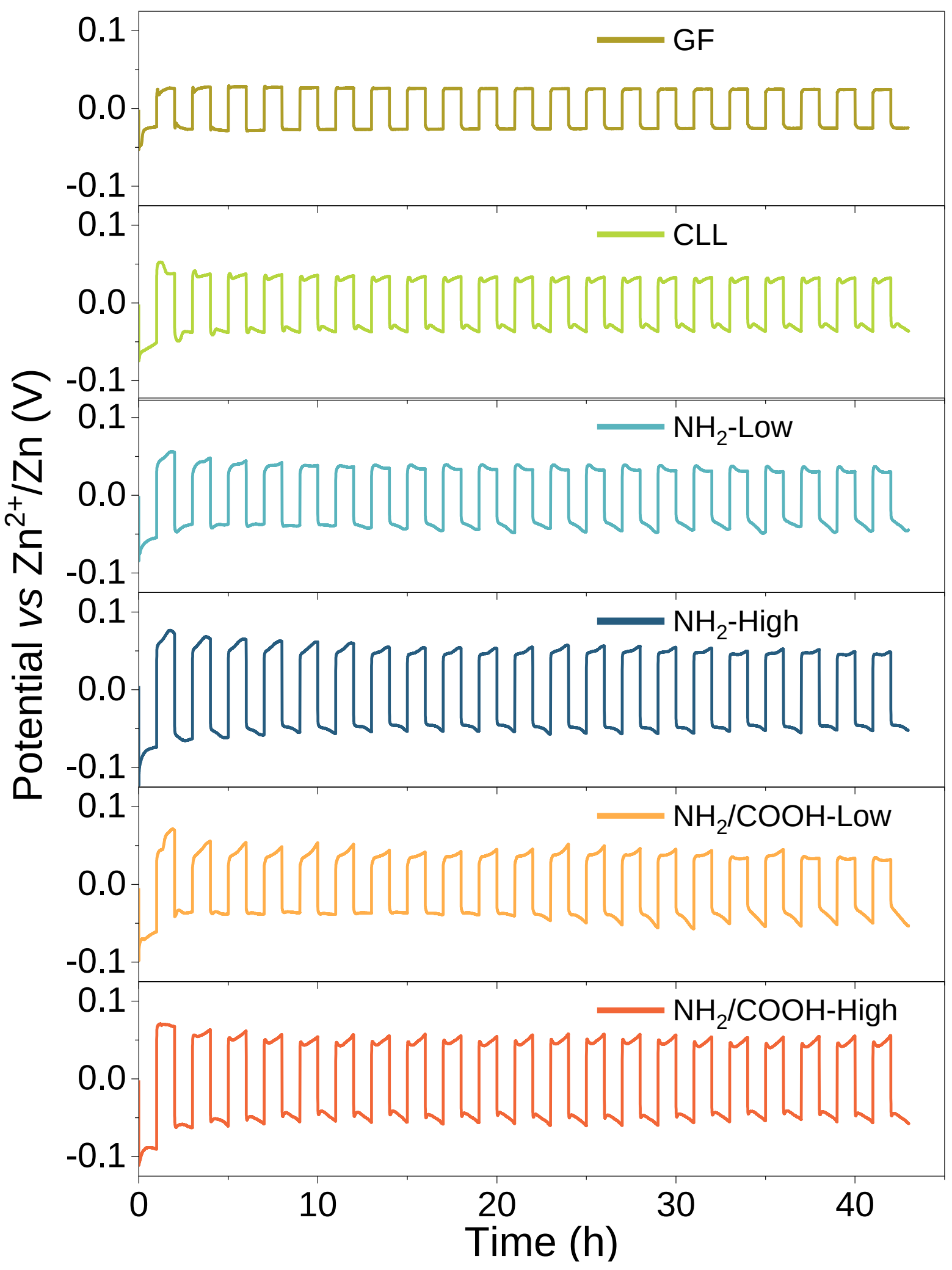

*Fig. S17: Plating and stripping profiles for 20 cycles with 1 mA/cm$^2$ and 1 mAh/cm$^2$ ending with a discharge.*

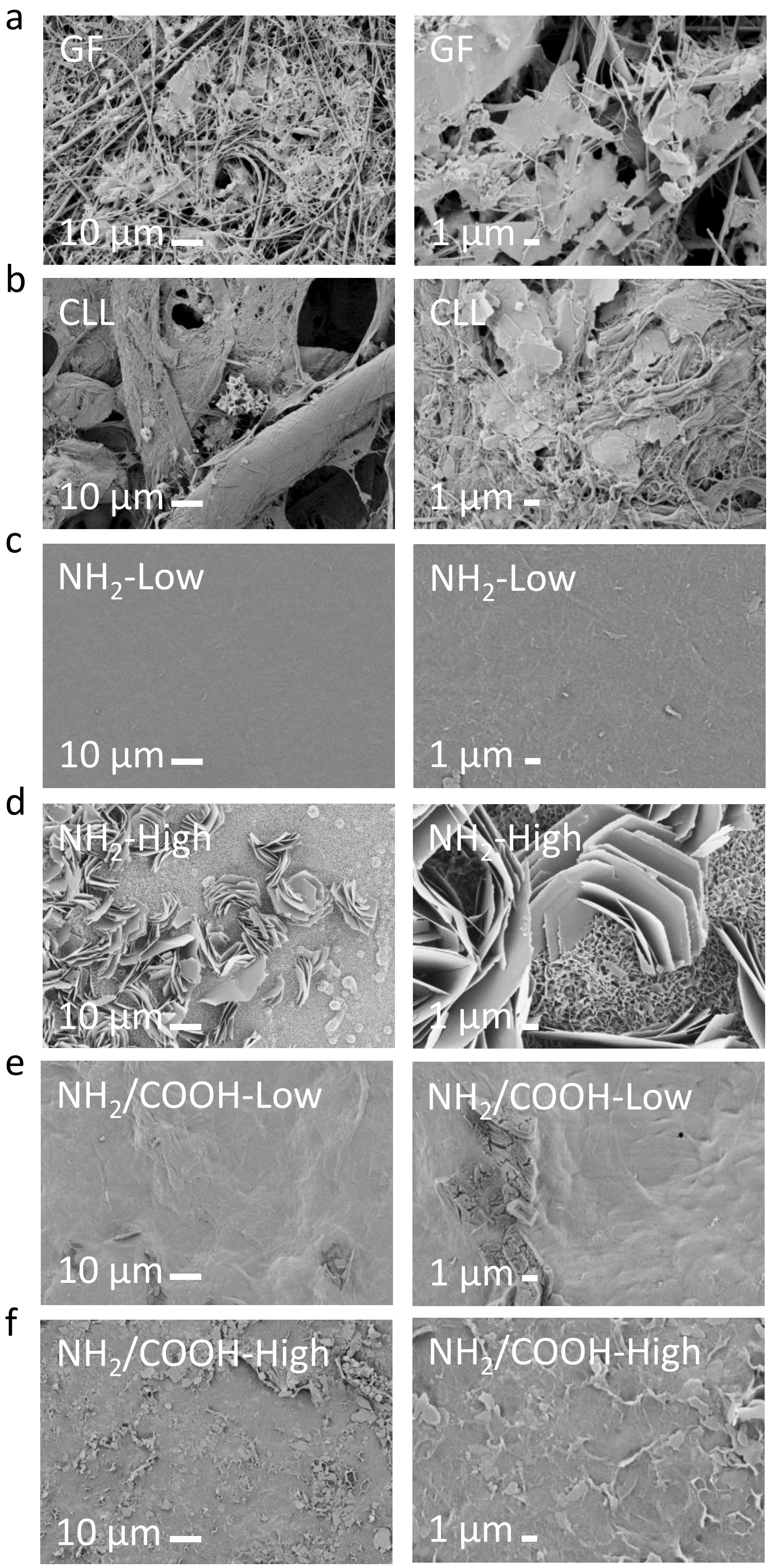

*Fig. S18: SEM images of recovered separators at different magnifications.*

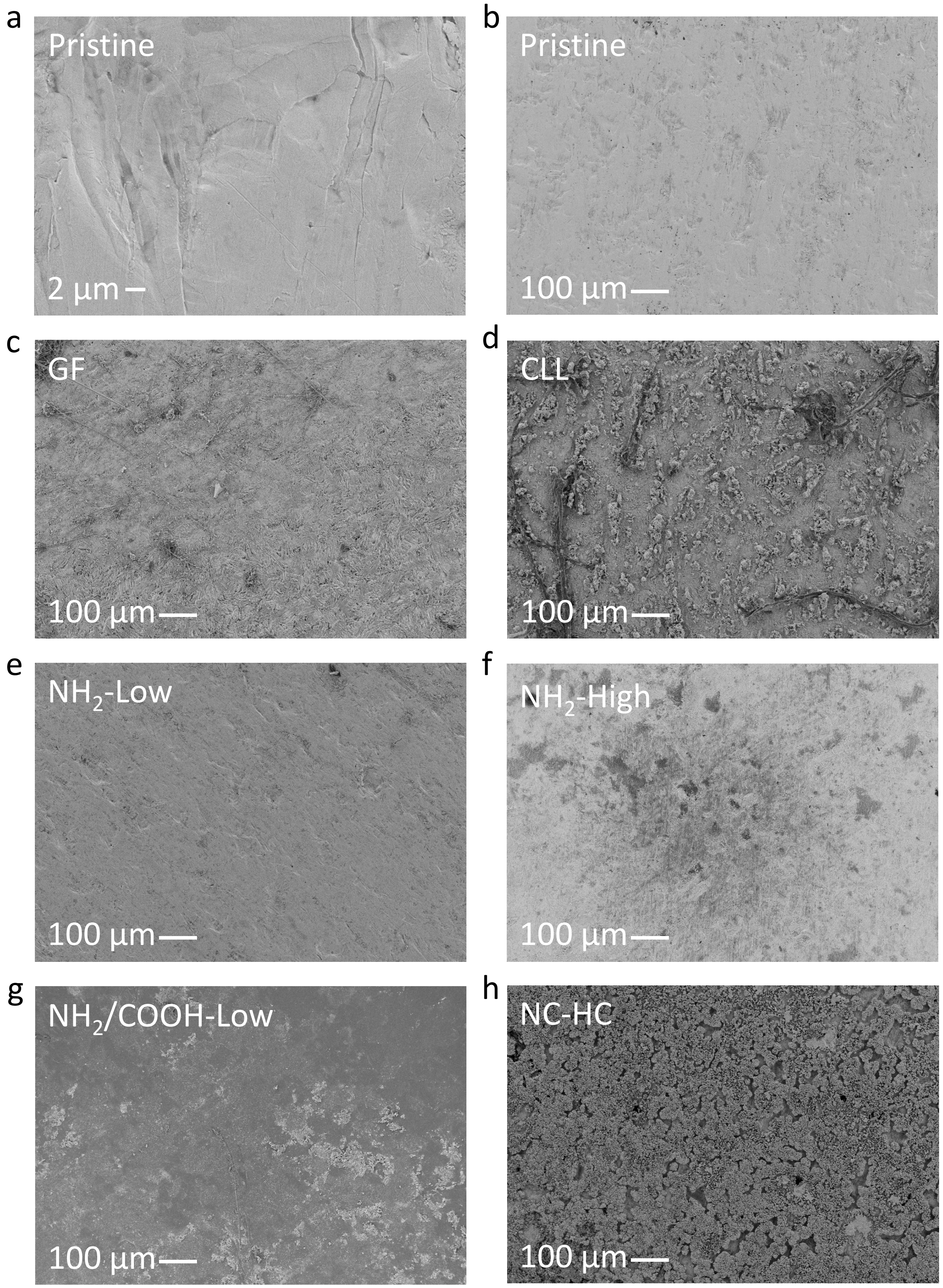

*Fig. 19: SEM images of the recovered Zn anodes at lower magnification.*

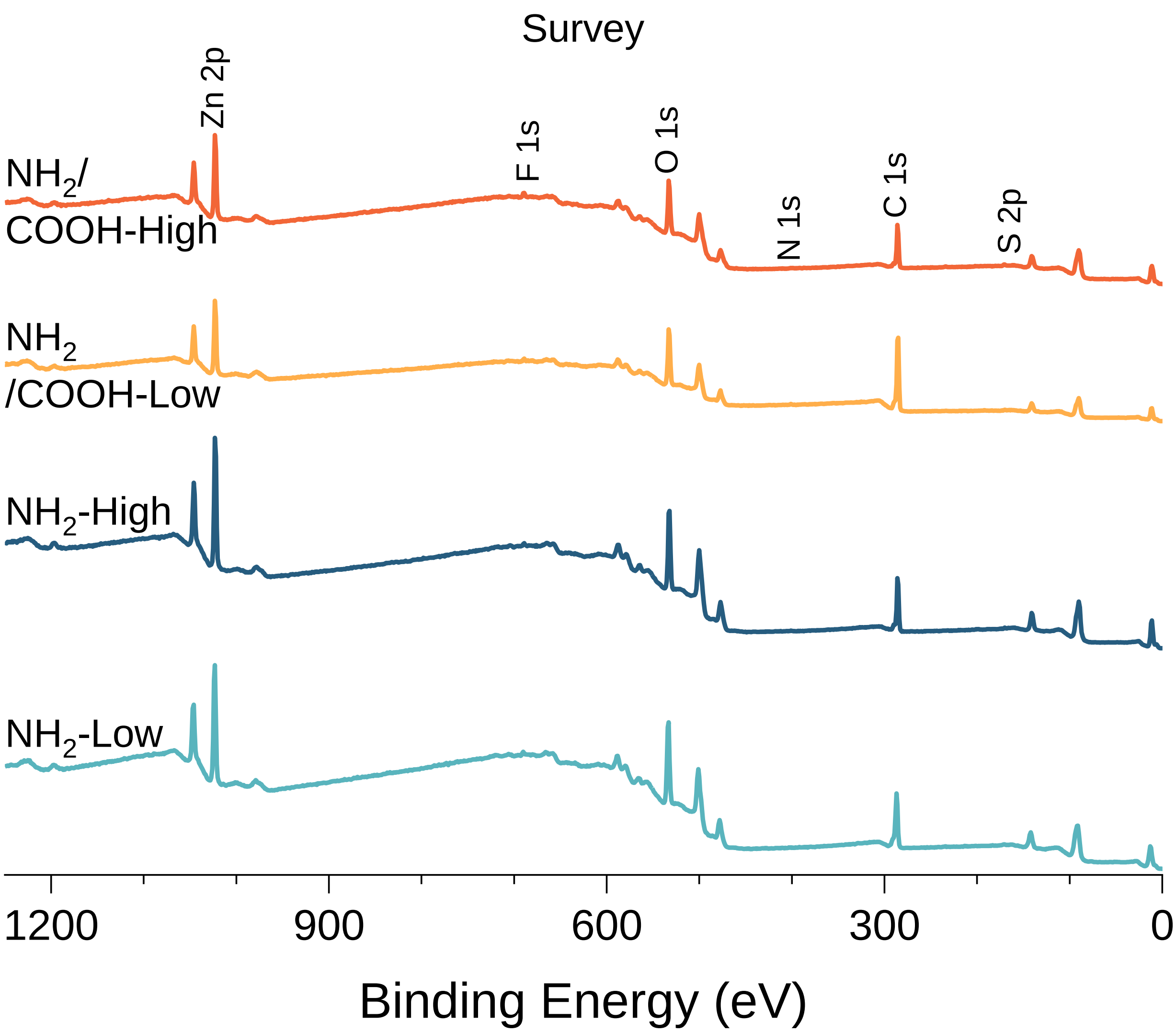

*Fig. 20: Survey of the XPS analysis.*

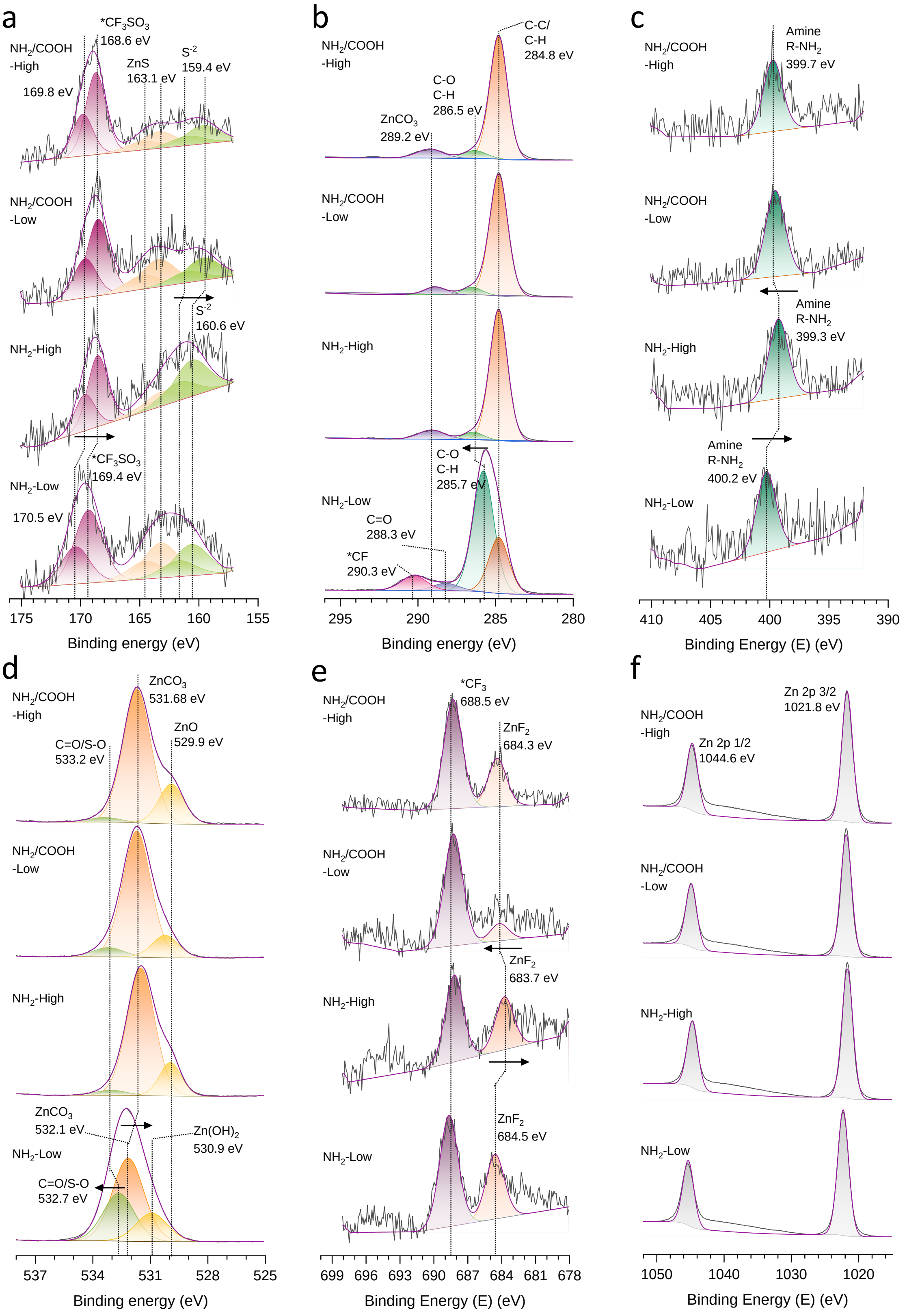

*Fig. S21: Elemental XPS analysis with the corresponding deconvolution. a) S 2p, b) C 1s, c) N 1s, d) O 1s, e) F 1s and f) Zn 2p spectra.*

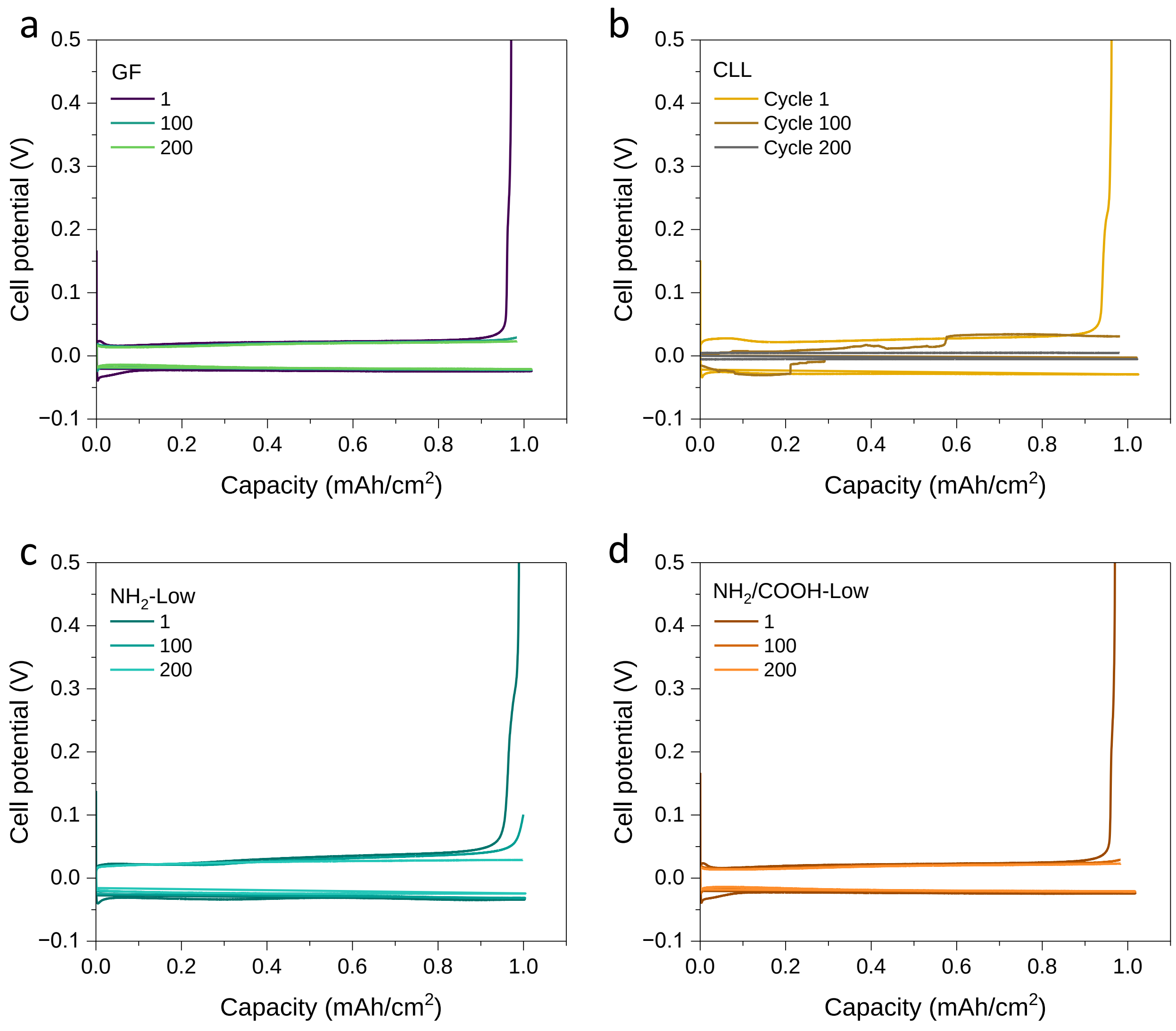

*Fig. 22: Profiles of CE with different separators utilizing a Zn‖Cu cell.*

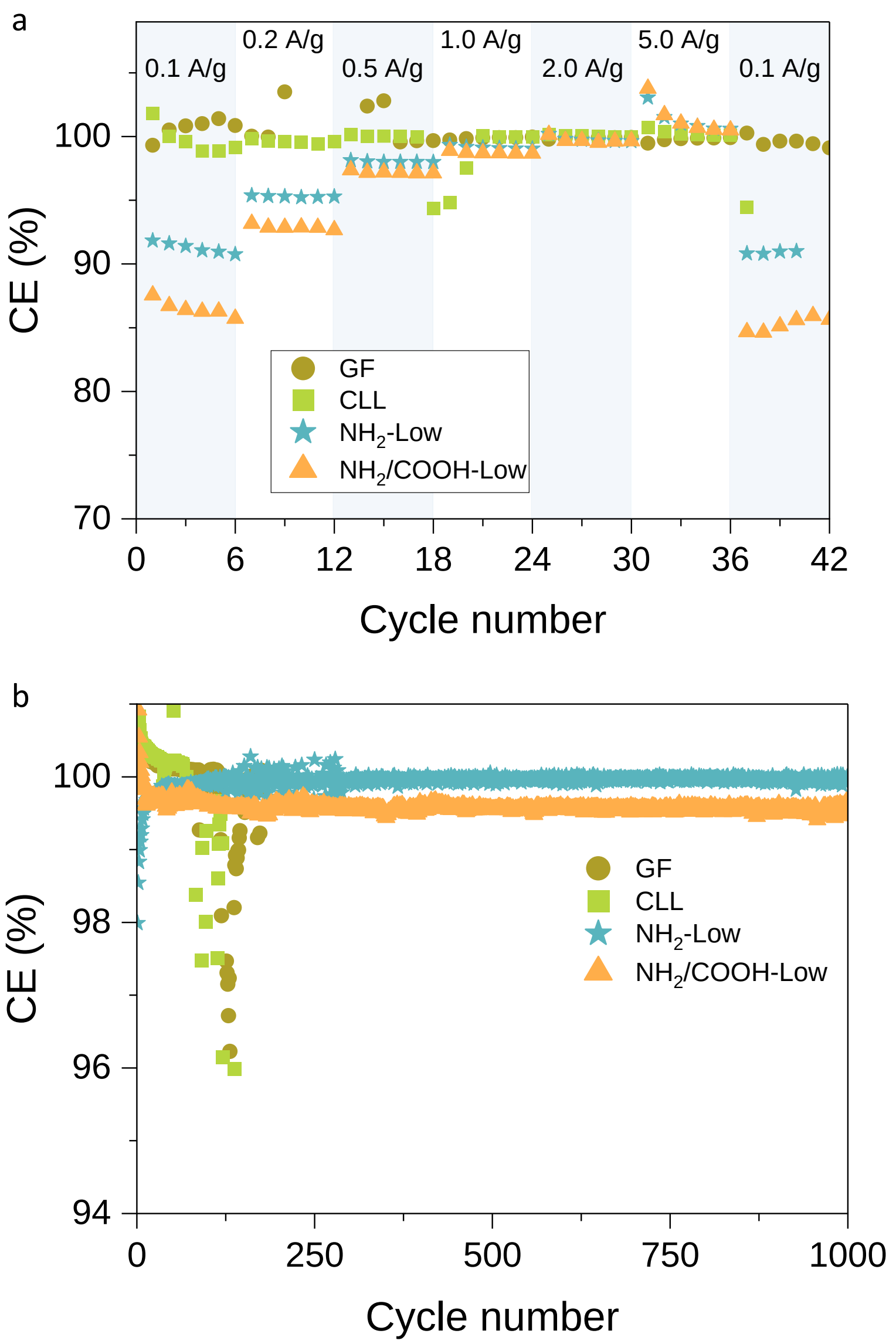

*Fig. S23: Coulombic efficiencies of the a) rate capability and b) long-term cycling at 2 A/g.*

## Tables S1-2

*Table 1: Titration results for functional group contents.*

| Sample | Primary amine groups (µmol/g) | Secondary amine groups (µmol/g) | Carboxylate groups (µmol/g) | Total functionality (µmol/g) |
|---|---|---|---|---|
| $NH_2$-Low | 780 | -- | -- | 780 |
| $NH_2$/COOH-Low | 660 | 120 | 120 | 900 |
| $NH_2$-High/ COOH-Low | 270 | 510 | 510 | 1290 |
| $NH_2$-High | 1740 | -- | -- | 1740 |
| $NH_2$/COOH-High | 1560 | 180 | 180 | 1920 |

*Table 2: Mechanical properties of the separators.*

| Sample | Tensile strength (MPa) | Young's modulus (MPa) |
|---|---|---|
| $NH_2$-Low | 6.3 ± 0.7 | 311.4 ± 55 |
| $NH_2$/COOH-Low | 15.2 ± 1.3 | 502.3 ± 132 |
| $NH_2$-High | 18.5 ± 1.1 | 323.5 ± 100 |
| $NH_2$/COOH-High | 15.2 ± 1.2 | 529.2 ± 66 |